\documentclass[%
 reprint,
superscriptaddress,
amsmath,amssymb,
 prl,
 aps,
]{revtex4-2}
\usepackage{graphicx} % Required for inserting images
\usepackage{float}
\usepackage{epstopdf}
\usepackage{subfig}
\usepackage{amsmath}
\usepackage{amssymb,amsthm}
\usepackage{indentfirst}
\usepackage[utf8]{inputenc}
\usepackage{centernot}
\usepackage{bm}
\usepackage{braket}
\usepackage{graphics}
\usepackage{amsfonts}
\usepackage{cmap}
\usepackage[dvipsnames]{xcolor}
\usepackage{amscd}

\usepackage{ragged2e}

\begin{document}

\title{Xenophobia based on a few attributes can impede society's cohesiveness}

\author{Alejandro Castro}
\affiliation{Group of Complex Systems and Statistical Physics, Department of
Theoretical Physics, University of Havana, Havana, Cuba}
\author{Tuan Minh Pham}
\affiliation{Dutch Institute for Emergent
Phenomena, Amsterdam 1090 GL, The Netherlands}
\affiliation{Institute of Physics, University of
Amsterdam, Science Park 904, The Netherlands}
\author{Ernesto Ortega}
\affiliation{Complexity Science Hub Vienna, Metternichgasse 8, 1030 Vienna, Austria}
\author{David Machado}
\email{\href{mailto:david.machado@uniroma1.it}{{\color{blue}david.machado@uniroma1.it} }}
\affiliation{Group of Complex Systems and Statistical Physics, Department of Theoretical Physics, University of Havana, Havana, Cuba}
\affiliation{Dipartimento di Fisica, Sapienza Università di Roma, P.le Aldo Moro 5, 00185 Rome, Italy}
\affiliation{CNR - Nanotec, unit\`a di Roma, 00185 Italy}

\begin{abstract}
  Xenophobic interactions play a role as important as homophilic ones in shaping many dynamical processes on social networks, such as opinion formation, social balance, or epidemic spreading.
  In this paper, we use belief propagation and Monte Carlo simulations on tree-like signed graphs to predict that a sufficient propensity to xenophobia can impede a consensus that would otherwise emerge via a phase transition. As the strength of xenophobic interactions and the rationality of individuals with respect to social stress decrease, this transition changes from continuous to discontinuous, with a strong dependence on the initial conditions. The size of the parameter region where consensus can be reached from any initial condition decays as a power-law function of the number of discussed topics. %The time to reach consensus  grows exponentially with the number of individuals if start from a random initialization.  The larger the number of topics discussed, the more prominent such a dependence on the initial conditions is.
\end{abstract}

\maketitle

Opinion dynamics underlie a variety of emergent social phenomena, including consensus \cite{castellano2009statistical, JUSUP20221, Redner2019}, polarization \cite{dimaggio1996polarized,fiorina2008political}, fragmentation \cite{granovetter1978threshold,gleeson2008cascades}, emergence of leaders \cite{bonabeau1995phase,mistry2015committed}, or cultural and norm formations \cite{axelrod1997dissemination,FRIEDKIN2001}. Therefore, extensive work has been devoted to studying how network structure determines opinion outcomes \cite{sood2005voter, Holme, Vazquez2008, Durrett2012, Baumann, Korbel2023,pham2025polarisation}. 

Interpersonal relationships often go beyond pairwise interactions because agents tend to form closed triangles under the so-called triadic closure effect \cite{newman2018networks}. In networks with both positive and negative interactions, there are two different types of triangles: those consisting of an odd number of negative edges and those with an even number of negative edges. According to Heider's social balance theory \cite{Heider, cartwright1956structural}, the former is called an unbalanced triad, and the latter a balanced triad. The theory posits that agents tend to minimize tension by reducing the number of unbalanced triads over time, resulting in a reorganization of their network \cite{minh2021, Gorki2020, Hao2024, Gallo2024, Kirkley2019, Szell, Leskovec, antal2005social, ANTAL2006, marvel2009, marvel2011continuous, Belaza, Hirotaka, Burda2022, Gorski2023, Oishi, Dekker, Malarz}, and, in particular, in the emergence of echo chambers \cite{Pham2020, Vendeville}.

Unlike these explanations for triangle formation using triadic relationships, a mechanism based solely on pairwise interactions was recently proposed by Pham \textit{et al.} \cite{Pham2022}. There, the interaction between any two agents is assumed to depend on their opinion similarity. Positively-linked agents tend to align due to homophily, while negatively-linked agents tend to disagree due to xenophobia (anti-homophily). Without explicit triadic interactions, this mechanism produces triangle patterns consistent with social network data and has been empirically validated \cite{Galesic2025}. However, the focus of \cite{Pham2022} was not on opinion formation, leaving unanswered an interesting question on how consensus is affected by the joint action of homophilic and xenophobic ties. 

This paper addresses this question by studying a system of agents interacting via both homophilic and xenophobic couplings on random sparse graphs. To this end, we apply belief propagation (BP), an exact method in tree-like graphs (i.e. those without short cycles), such as Erdős–Rényi (ER) or random regular (RR) \cite{Kabashima_1998,yedidia2005constructing, pearl1988probabilistic, kschischang2001factor,Yedidia2000,yedidia2003understanding}, which can also be extended to networks with short loops \cite{Rizzo2010, Metz2011, bolle2013spectra, Kirkley, Cantwell2023, WangPRL, Guzman, PhamPRE2024}. In doing so, we can identify the conditions for reaching consensus via a disorder-to-order transition. As expected, when the number of discussed issues (denoted here as $g$) increases, consensus becomes harder to find. Although the existence of a maximum value of $g$ was previously suggested \cite{Deffuant,castellano2009statistical}, in our case, the feasibility of consensus decreases as a power-law of the number of discussed issues. Since this law is scale-free,  there is no characteristic value of $g$ that guarantees consensus.

{\bf \textit{ Model.}} We consider a system of $N$ individuals connected via a network $\mathcal{G}(V, E)$, where $V$ denotes the set of vertices and $E$ the set of edges. This network is associated with an adjacency matrix $\mathbf{A}$ with $A_{ij} = 1$ for $e_{ij}\in E$ and $A_{ij}=0$ otherwise. Each node $i\in V$ has $k_i$ neighbors and opinions on $g$ different ``issues", forming a feature vector $\mathbf{s}_i$ of dimension $g$ with $s^\mu_i \in \{-1, 1\}$, for $\mu = 1, 2, \cdots, g$. This captures, for instance, an election process in which a candidate is to be selected based on multiple topics, such as abortion, public \textit{vs.} private health care, climate change, gun control, etc. 

The model assumes that individuals distribute their cognitive attention evenly across all topics. Although real-world agents typically experience bounded rationality and finite attention spans, this deliberate simplification allows us to isolate the purely structural effects of opinion dimensionality.

The interactions co-evolve dynamically with opinions: agents adjust their couplings $J_{i,j} = {\rm sign}(\mathbf{s}_i \cdot \mathbf{s}_j)$ based on the opinion similarity, while $J_{i,j}$, in turn, affects the update of opinions through social pressure specified by a Hamiltonian. Specifically, the \emph{local} Hamiltonian of a node $i$ having both  homophilic and xenophobic interactions reads: 
\begin{equation}
  H^{(i)}({\bf s}_i, {\bf s}_{\partial i}) = \! -\frac{\alpha}{g} \!\! \sum_{j:J_{i,j}=1} \!\!\! \mathbf{s}_i \cdot \mathbf{s}_j + \frac{1- \alpha}{g} \!\!\! \sum_{j:J_{i,j}=-1} \!\!\! \mathbf{s}_i \cdot \mathbf{s}_j
  \label{hamitonian}
\end{equation}
where $\alpha$ is a parameter termed affinity that controls the propensity to focus on homophilic interactions, and $(1-\alpha)$ that to focus on xenophobic ones. As $\alpha$ decreases, the effect of xenophobic interaction becomes stronger. This local Hamiltonian (social stress) accounts for the influence that an agent $i$ receives from interactions with its friends through the first term and from those with its enemies through the second term. Monte Carlo simulations of this model consist of a large number of steps, at each of which we select one agent and its attribute, both at random. Then, we propose a new attribute that is accepted or rejected using the Metropolis scheme \cite{Metropolis1953} at a given temperature $T$. In a social context, $T$ characterizes the average volatility of individuals: the higher $T$, the more likely an agent is to update its opinion, regardless of whether that update reduces social stress \cite{Bahr1998}. Since individuals are assumed to update opinions and social links on the same timescale, once the opinion has been updated, we recompute their couplings $J_{i,j} = {\rm sign}(\mathbf{s}_i \cdot \mathbf{s}_j)$ before continuing the next step of the opinion dynamics.

 {\bf \textit{Belief Propagation Equations.}} The \emph{marginal} probability $P({\bf s}_i)$ of $i$'s opinion vector is given by
\begin{equation}
  P(\mathbf{s}_i) \propto \sum_{\mathbf{s}_{\partial i}} e^{\beta H^{(i)}({\bf s}_i, {\bf s}_{\partial i})} \prod_{j \in \partial i} m_{j \rightarrow i}(\mathbf{s}_j) \label{eq:P_local}
\end{equation}
where  $\partial i$ denotes the set of neighbors of $i$ and $\beta = T^{-1}$ represents the rationality of individuals with respect to social stress. The factors in the product are the marginal probabilities (called messages) of the opinion ${\bf s}_j$, where $j \in \partial i$, in a modified graph where the edge $(i,j)$ is removed. Since the graph is tree-like, in a scenario with finite correlation length, the removal of $(i,j)$ effectively disconnects the nodes $i$ and $j$ and makes $m_{j \rightarrow i}(\mathbf{s}_j)$ independent of ${\bf s}_i$. As derived in Appendix A, these messages can be computed recursively using the following  rule:
 \begin{equation}
 \begin{aligned}
  m_{j \rightarrow i}({\bf s}_j) \! \propto \!\!\! \prod_{k \in \partial j \setminus i} \sum_{{\bf s}_k} m_{k\rightarrow j}({\bf s}_k) \; e^{\frac{\beta}{2g} ({\bf s}_k \cdot {\bf s}_j) [ 2 \alpha - 1 + {\rm sgn} ({\bf s}_k \cdot {\bf s}_j) ] } \label{eq:BP_rule}
  \end{aligned}
\end{equation}
This set of equations can be solved iteratively to find the messages (if a stable point of this iterative scheme exists). Subsequently, using these final BP's messages, we can compute $P(\mathbf{s}_i)$ using Eq. (\ref{eq:P_local}).
To measure the level of agreement, we define the magnetization as follows: 
\begin{equation}
m = \Big|\Big| \frac{1}{N} \sum_{i} \sum_{{\bf s}_i} P({\bf s}_i) \,\mathbf{s}_i \Big| {\Big|}_{2}
\end{equation}
where $| {|} \cdot | {|}_{2}$ is the Euclidean norm. Note that $m=0$ if the opinions ${\bf s}_i$ have no preferred direction, while $m>0$ if the opinions are at least partially aligned along a given direction. When all people share the same opinions,  a global consensus is achieved, so $m=1$. At each temperature $T=\beta^{-1}$, there exists a critical value $\alpha_c(T)$ such that $m=0$ (paramagnetic state) for $\alpha < \alpha_c(T)$, and $m>0$ (ferromagnetic state) for $\alpha > \alpha_c(T)$. Therefore, $\alpha_c(T)$ marks a disorder-to-order transition. To find $\alpha_c(T)$ (with a precision equal to $d\alpha$), we start at some $\alpha_0 < \alpha_c(T)$ and measure $m(\alpha_0)$. As long as $m(\alpha_0)=0$, we set $\alpha_1=\alpha_0 + d\alpha$ and repeat the process until $m(\alpha_i)>0$.

Since BP does not converge for $\alpha < 0.5$, we focus only on $\alpha \geq 0.5$. This lack of convergence usually indicates the existence of a frustrated phase, due to the high non-convexity of the underlying solution space with multiple local minima \cite{MezardMontanariBook2009}. 
%In fact, for the simple case with only one issue to discuss ($g=1$), when $\alpha<0.5$ we have nothing more than the well-known antiferromagnetic model in random graphs.
Intuitively, at $\alpha <0.5$, the system is more likely to be driven by the agents' tendency to separate their opinions from those of their neighbors. At low temperatures and due to the presence of long loops in the graph, individuals face inconsistencies in separating themselves from the rest, and a frustrated phase arises.

\begin{figure}[t]
\includegraphics[width=\linewidth]{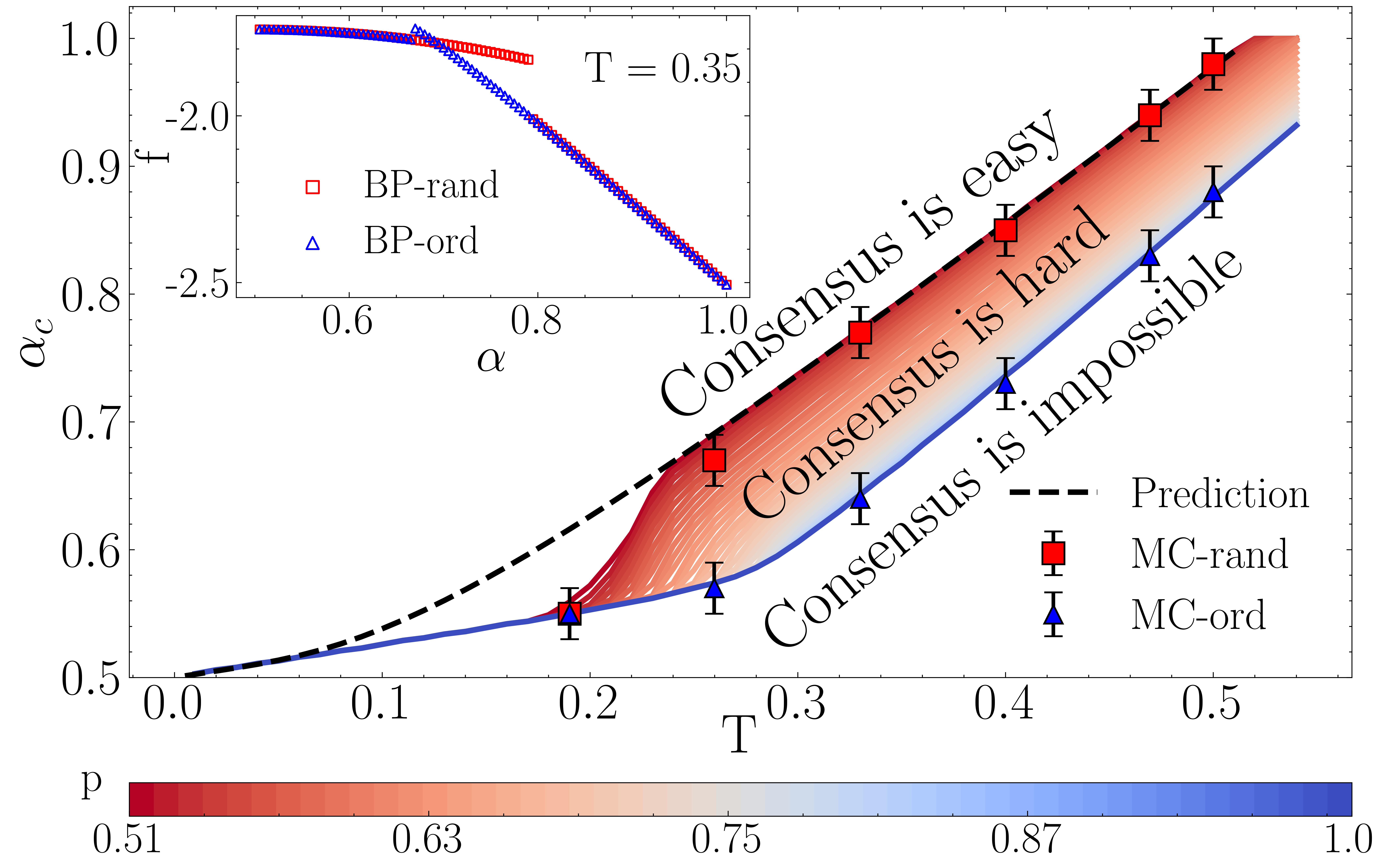}
 \caption{\justifying Phase diagram of the model in random regular graphs with $g=c=5$. The figure includes BP's predictions (solid lines), the results of Monte Carlo simulations (squares and triangles), and our analytic computation of the stability of the paramagnetic solution (dashed lines). Different colors represent BP's predictions for different initial conditions obtained using a binomial distribution (see Eq. \ref{eq:init_binomial}) to assign the initial value of the messages. Monte Carlo simulations are performed over systems with $N=10^{4}$ individuals. When $p=0.5$, the initial state has zero magnetization (no alignment of the opinions). When $p=1.0$, the initial state is completely aligned ($m=1$). BP's transition lines have an uncertainty of order $d\alpha=0.001$, not shown since it is of the size of the lines themselves. The inserted graphic shows the two branches of BP's free energy density for different $\alpha$ at $T = 0.35$. The blue triangles are obtained with completely aligned initial conditions, and the red squares with random initial conditions.} 
 \label{fig:mc_vs_bp_RRG}
\end{figure}

{\bf \textit{Results.}} Fig. \ref{fig:mc_vs_bp_RRG} shows the dependence of consensus on the level of affinity in random regular graphs. As $\alpha$ increases beyond $\alpha_c(T)$, the system goes from disagreement to consensus. The higher the temperature $T$, the larger $\alpha_c(T)$. In a random regular graph, given that all links are equivalent, we can reduce Eq. \eqref{eq:BP_rule} to a single equation for one message by initializing all messages $m_{j \to i}({\bf s}_j)$ with the same (random or aligned) value. This further allows us to study the dependence of the BP solution on the initial conditions. Specifically, assuming $m_{j \to i}({\bf s}_j)$ is independent of the edge $(j,i)$ but depends only on the number of positive components of $j$'s opinion vector $n = \sum_{\mu=1}^{g} \delta_{s_j^{\mu}, 1}$, we can set the initial messages using a binomial form:
\begin{equation}
  m(n) = \binom{g}{n} \, p^{n} \, (1-p)^{g-n} \label{eq:init_binomial}
\end{equation}
where $p \in [0.51, 1]$ is a parameter that can be continuously varied from completely random configurations ($p=0.5$) to a fully aligned one ($p=1$). 

In Fig. \ref{fig:mc_vs_bp_RRG}  we use colored lines to depict the transitions obtained by BP for different initial conditions, each given by a particular value of $p$. The disorder-to-order transition obtained with $p=0.51$ is marked by the darkest tone of red, while that with $p=1$ is the darkest blue. In the intermediate region, the system has a peculiar behavior. There, although consensus exists as an attractor for the dynamics, it would take a time that grows exponentially with the number $N$ of agents for a system starting from a random configuration to reach consensus. The larger the number of topics discussed, the more pronounced this dependence on initial conditions becomes.

The phase diagram in Fig. \ref{fig:mc_vs_bp_RRG} is divided into three distinct regions. Above the darkest red line, it is easy to reach a consensus from any initial condition. Below the darkest blue line, it is impossible to achieve consensus, as it is not an attractor of the dynamics. There is a region in between, where it is hard but possible to attain consensus due to a strong dependence on the initial conditions.

The lines generated using $p\in(0.51, 1)$ continuously cover the area between these two extremes. Nevertheless, we observe that a small perturbation ($p \approx 0.63$) already shifts the transition far away from the results obtained with $p\sim 0.5$. On the other hand, the results for $p>0.75$ remain very close to what is obtained with $p=1$. This shows how, at a fixed temperature, a small perturbation in the right direction can facilitate consensus.  In Fig. \ref{fig:mc_vs_bp_RRG} we also verify that, regardless of the value of $p$, there exists a critical point $(T_c, \alpha_c(T_c))$ such that, for $T>T_c$, there are two different transitions, depending on whether we start from random or aligned initial conditions. Exactly at $(T_c, \alpha_c(T_c))$, all the colored lines merge into a single one, which remains unique for all $T<T_c$. 

To better understand the phase transition, we show the free energy density of the system, obtained using BP, in the inset of Fig. \ref{fig:mc_vs_bp_RRG}.  At $T> T_c$, the free energy exhibits two distinct branches, confirming a first-order transition, typically accompanied by a strong dependence on initial conditions. The \emph{true} free energy is always the minimum among the two branches and has a discontinuity in its derivative at the crossing point of the branches, where the actual thermodynamic transition takes place. On the other hand, at $T < T_c$ the free energy has only one branch and no observable discontinuities in its derivative (see Appendix F). In other words, the nature of the phase transition changes from discontinuous at high $T$ to continuous at low $T$, with the point $(T_c, \alpha_c(T_c))$ separating these two regimes. 

We next compare BP's predictions with Monte Carlo simulations for $N=10^{4}$ and two different initial conditions: i) globally aligned opinions (blue points) or ii) iid random opinions at each node (red points). In Appendix D, we demonstrate a strong dependence of the results on the initial conditions. Remarkably, the critical values of $\alpha_c$ obtained by BP with $p=1.0$ and $p=0.5$, respectively, are in excellent agreement with simulations. 

We complete our analysis of the phase transition by \emph{approximately} computing the transition from disorder to order. Our prediction is represented by a dashed black line in Fig. \eqref{fig:mc_vs_bp_RRG} and is obtained from the stability analysis of BP's disordered solution. Specifically, we apply a small perturbation to the paramagnetic fixed point and expand Eq. \ref{eq:BP_rule} to the first order in powers of the perturbation. We then study whether the perturbed messages will return to the paramagnetic fixed-point solution, or, instead, will depart from it. In the latter case, the fixed point is no longer stable. The details are in Appendix B. 

Expanding to the first order in the perturbation is just an upper bound, since subsequent powers could destroy the stability of this fixed point. It is not surprising that this works well only with first-order phase transitions, for large $\alpha$ and $T$. Our methods can be directly applied to other graph ensembles like Erdős-Rényi, as we will show later, and their numerical evaluation is considerably faster than solving the BP or Monte Carlo simulations. 

\begin{figure}[t]
\includegraphics[width=\linewidth]{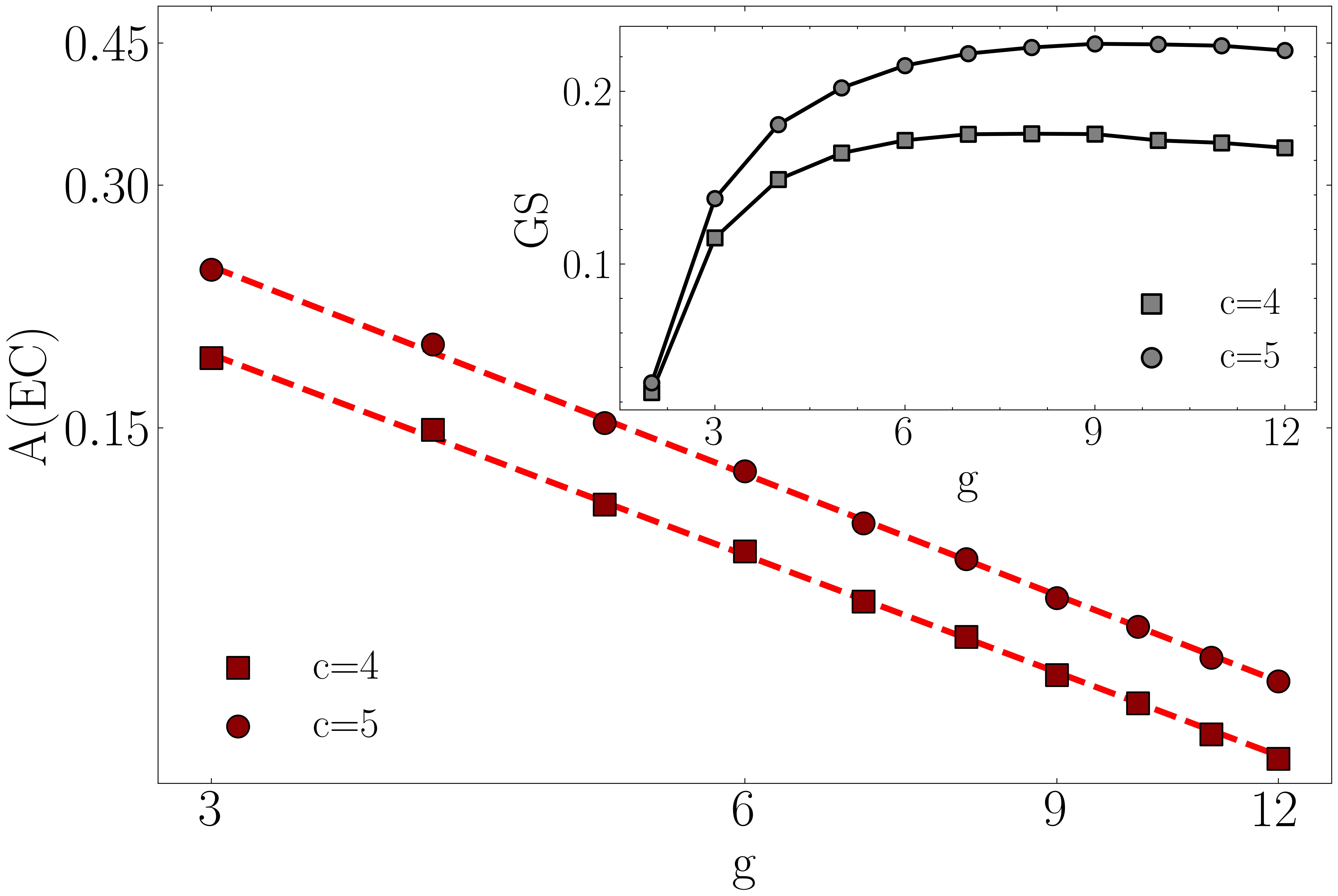} 
  \caption{\justifying Area of easy consensus A(EC) \textit{vs.} the length $g$ of the opinion vectors on a logarithmic scale in random regular graphs with fixed connectivity $c=4$ (square) and $c=5$ (circle). Red dashed lines are power-law fits to the data points: $A(EC) = \Phi(c) \,g^{-\gamma(c)}$. The optimal parameters depend on the connectivity. They are: $\Phi(4)=0.459(8)$, $\Phi(5)=0.61(1)$, and $\gamma(4)=0.83(1)$, $\gamma(5)=0.85(1)$. The inserted graphic shows the gap's size GS defined in Eq. \eqref{eq:gap_size} as a function of $g$.}
\label{fig:gap_size_RRG}
\end{figure}

At each temperature,  there exists a finite gap $\alpha_c^{\text{rand}}(T)-\alpha_c^{\text{ord}}(T) >0$ due to the existence of two diffe\-rent transitions, one for random ($p=0.51$) and another for aligned ($p=1$) initial condition. We can quantify the difficulty of reaching a consensus, even when it is possible, by introducing two definitions. Let A(EC) (area of easy consensus) be the area above the darkest red line, where consensus is easy, and A(PC) (area of possible consensus) be the entire area above the darkest blue line, where consensus is possible.  Fig. \ref{fig:gap_size_RRG} shows that discussing a large number of issues makes it harder to reach consensus. Indeed, the area A(EC) goes to zero when $g$ increases. Its behavior for each connectivity $c$ is well described by a power-law decay $A(EC) = \Phi(c) g^{-\gamma(c)}$, whose parameters depend on $c$. For both $c=4$ and $c=5$, the exponent $\gamma$ is smaller than one, signaling the scale-free nature of the decay.  Remarkably, there are no characteristic lengths.

The difference $\text{A(PC)}-\text{A(EC)}$  gives the area where consensus is possible, but hard to find if one starts with random initial conditions. Normalizing this difference by A(PC),  we have the following measure of the gap's size:
\begin{equation} \text{GS} = \frac{\text{A(PC)}-\text{A(EC)}}{\text{A(PC)}} \label{eq:gap_size}
\end{equation}
The behavior of this GS is shown in the inserted graphic in Fig. \ref{fig:gap_size_RRG}. This space where consensus is hard to find grows
with the number of topics until it stabilizes above $g \sim 8$. For completeness, we show that A(PC) does not follow a power-law behavior in Appendix E.

\begin{figure}[t]
\includegraphics[width=\linewidth]{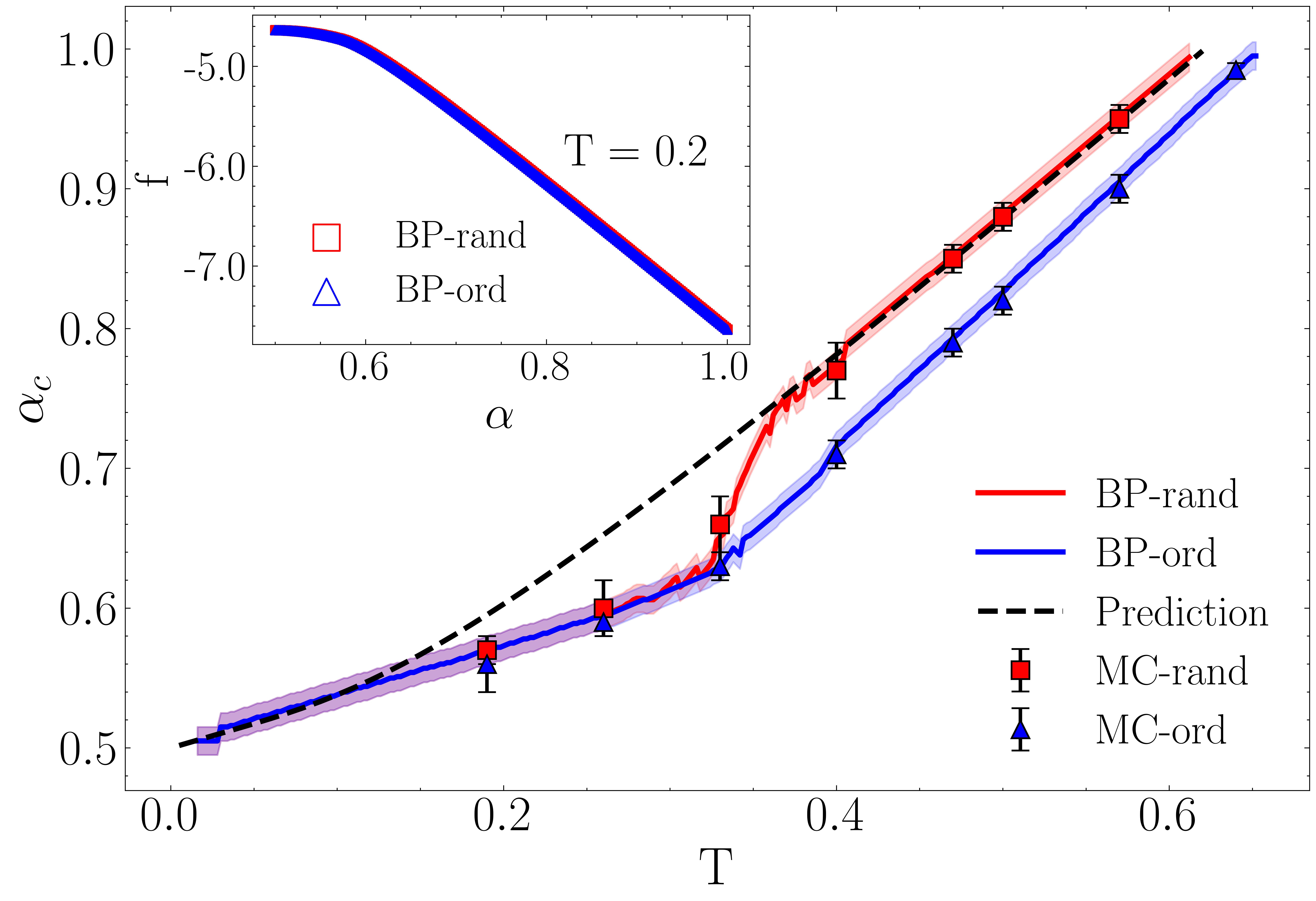}
                \caption{ \justifying Phase diagram of the model in Erdős-Rényi graphs. The figure includes BP's predictions (solid lines), the results of Monte Carlo simulations (squares and triangles), and our approximation of the stability of the paramagnetic solution (dashed lines). Monte Carlo simulations were carried out for systems with $N = 10^5$ individuals, $g=3$, and average connectivity $\kappa=3$. The red points and lines are obtained by initializing all the opinions uniformly at random. The blue points and lines are instead the result of initializing the opinions aligned toward a given direction. The colored shadows represent the uncertainty in the determination of BP's transition lines, which is of order $d\alpha=0.01$ in this case. The inserted graphic shows BP's free energy density as a function of $\alpha$ at $T = 0.2$. The blue triangles are obtained with completely aligned initial conditions, and the red squares with random initial conditions.}
  \label{fig:ER_phase_diagram}
\end{figure}

Finally, in Fig. \ref{fig:ER_phase_diagram}  we confirm the validity of our results, including the approximate computation for the stability of the paramagnetic solution, in Erdős-Rényi random graphs.  The phenomenology is essentially the same as in random regular graphs. The inserted graphic shows that for low $T < T_c$ the free energy is continuously differentiable. To complete this analogy, we include a figure for the free energy at high $T>T_c$ in Appendix F.

There are some numerical challenges to obtaining accurate results for Erdős-Rényi graphs. As the connectivity is not uniform, we used a standard population dynamics algorithm \cite{MezardParisi2001, mozeika2009dynamical} to sample the distribution of BP's messages. Furthermore, Monte Carlo simulations also suffer from large finite-size effects in this case. Nevertheless, we find good agreement between BP's predictions and Monte Carlo simulations.

In summary, we demonstrate the impact of xenophobic interactions on opinion formation in sparse network topologies. The results deliver a general message: under the joint effect of homophily and xenophobia, a society
with the ability of a coevolutionary dynamics of opinion-
and link formation must be expected to have a phase diagram structure as that presented in figure \ref{fig:mc_vs_bp_RRG}.  Moreover, we find that consensus can emerge via both discontinuous and continuous phase transitions. These transitions are remarkably robust under changes in the network topology.  Such a mixed-order phase transition has recently been observed in other opinion dynamics models \cite{jan2025, Gorski2025,Noudehi}.  

We can also use BP to obtain other observables, such as the total level of social stress in the whole society and the fraction of positive links. Preliminary results for these observables are presented in Appendix C, highlighting the need for further investigation to understand their non-trivial behaviors. 

Tree-like networks that we consider do not have many triangles. Therefore, it would be interesting to study the effect of xenophobia in networks having short loops using generalized BP \cite{Lage-Castellanos2008}.  We can also study other related models, such as the Axelrod model \cite{Pedraza} or the integration of structural balance and status hierarchies \cite{Gorski2025}.

\pagebreak

\textit{Acknowledgement.} We thank Alejandro Lage Castellanos for his useful comments. Tuan Pham was supported in part
by the Dutch Institute for Emergent Phenomena  cluster
at the University of Amsterdam under the Research Priority Area
Emergent Phenomena in Society: Polarisation, Segregation and
Inequality. This study was conducted using the DARIAH HPC cluster at CNR-NANOTEC in Lecce, funded by the "MUR PON Ricerca e Innovazione 2014-2020" project, code PIR01\_00022.

\bibliographystyle{unsrt}
\bibliography{biblio} % Replace "references" with the name of your bibliography file

\begin{thebibliography}{10}

\bibitem{castellano2009statistical}
Claudio Castellano, Santo Fortunato, and Vittorio Loreto.
\newblock Statistical physics of social dynamics.
\newblock {\em Reviews of Modern Physics}, 81(2):591--646, 2009.

\bibitem{JUSUP20221}
Marko Jusup, Petter Holme, Kiyoshi Kanazawa, Misako Takayasu, Ivan Romić, Zhen
  Wang, Sunčana Geček, Tomislav Lipić, Boris Podobnik, Lin Wang, Wei Luo,
  Tin Klanjšček, Jingfang Fan, Stefano Boccaletti, and Matjaž Perc.
\newblock Social physics.
\newblock {\em Physics Reports}, 948:1--148, 2022.
\newblock Social physics.

\bibitem{Redner2019}
Sidney Redner.
\newblock Reality-inspired voter models: A mini-review.
\newblock {\em Comptes Rendus Physique}, 20:275--292, 2019.

\bibitem{dimaggio1996polarized}
Paul DiMaggio, John Evans, and Bethany Bryson.
\newblock Have american's social attitudes become more polarized?
\newblock {\em American Journal of Sociology}, 102(3):690--755, 1996.

\bibitem{fiorina2008political}
Morris~P Fiorina and Samuel~J Abrams.
\newblock Political polarization in the american public.
\newblock {\em Annual Review of Political Science}, 11:563--588, 2008.

\bibitem{granovetter1978threshold}
Mark Granovetter.
\newblock Threshold models of collective behavior.
\newblock {\em American Journal of Sociology}, 83(6):1420--1443, 1978.

\bibitem{gleeson2008cascades}
James~P Gleeson.
\newblock Cascades on correlated and modular random networks.
\newblock {\em Physical Review E}, 77(4):046117, 2008.

\bibitem{bonabeau1995phase}
Eric Bonabeau, Guy Theraulaz, and Jean-Louis Deneubourg.
\newblock Phase diagram of a model of self-organizing hierarchical systems: The
  case of the social insects.
\newblock {\em Physical Review Letters}, 75(8):1339--1342, 1995.

\bibitem{mistry2015committed}
Dina Mistry, Qian Zhang, Nicola Perra, and Andrea Baronchelli.
\newblock Committed activists and the resilience of crowd dynamics.
\newblock {\em Physical Review E}, 92(4):042805, 2015.

\bibitem{axelrod1997dissemination}
Robert Axelrod.
\newblock The dissemination of culture: A model with local convergence and
  global polarization.
\newblock {\em Journal of Conflict Resolution}, 41(2):203--226, 1997.

\bibitem{FRIEDKIN2001}
Noah~E. Friedkin.
\newblock Norm formation in social influence networks.
\newblock {\em Social Networks}, 23(3):167--189, 2001.

\bibitem{sood2005voter}
V~Sood and S~Redner.
\newblock Voter model on heterogeneous graphs.
\newblock {\em Physical Review Letters}, 94(17):178701, 2005.

\bibitem{Holme}
Petter Holme and M.~E.~J. Newman.
\newblock Nonequilibrium phase transition in the coevolution of networks and
  opinions.
\newblock {\em Phys. Rev. E}, 74:056108, 2006.

\bibitem{Vazquez2008}
Federico Vazquez, V\'{\i}ctor~M. Egu\'{\i}luz, and Maxi~San Miguel.
\newblock Generic absorbing transition in coevolution dynamics.
\newblock {\em Phys. Rev. Lett.}, 100:108702, 2008.

\bibitem{Durrett2012}
Richard Durrett, James~P. Gleeson, Alun~L. Lloyd, Peter~J. Mucha, Feng Shi,
  David Sivakoff, Joshua E.~S. Socolar, and Chris Varghese.
\newblock Graph fission in an evolving voter model.
\newblock {\em Proceedings of the National Academy of Sciences},
  109(10):3682--3687, 2012.

\bibitem{Baumann}
Fabian Baumann, Philipp Lorenz-Spreen, Igor~M. Sokolov, and Michele Starnini.
\newblock Modeling echo chambers and polarization dynamics in social networks.
\newblock {\em Phys. Rev. Lett.}, 124:048301, 2020.

\bibitem{Korbel2023}
Jan Korbel, Simon~D. Lindner, Tuan~Minh Pham, Rudolf Hanel, and Stefan Thurner.
\newblock Homophily-based social group formation in a spin glass self-assembly
  framework.
\newblock {\em Phys. Rev. Lett.}, 130:057401, 2023.

\bibitem{pham2025polarisation}
Tuan Pham, Sidney Redner, Lourens Waldorp, Jay Armas, and Han L.~J. van~der
  Maas.
\newblock Polarisation in increasingly connected societies.
\newblock {\em arXiv}, 2503.24098, 2025.

\bibitem{newman2018networks}
Mark Newman.
\newblock {\em Networks}.
\newblock Oxford university press, 2018.

\bibitem{Heider}
Fritz Heider.
\newblock Attitudes and cognitive organization.
\newblock {\em J. Psychol.}, 21(1):107--112, 1946.

\bibitem{cartwright1956structural}
Dorwin Cartwright and Frank Harary.
\newblock Structural balance: A generalization of {H}eider's theory.
\newblock {\em Psychological Review}, 63(5):277--293, 1956.

\bibitem{minh2021}
Tuan Minh~Pham, Andrew~C. Alexander, Jan Korbel, Rudolf Hanel, and Stefan
  Thurner.
\newblock Balance and fragmentation in societies with homophily and social
  balance.
\newblock {\em Scientific Reports}, 11:17188, 2021.

\bibitem{Gorki2020}
Piotr~J. G\'orski, Klavdiya Bochenina, Janusz~A. Ho\l{}yst, and Raissa~M.
  D'Souza.
\newblock Homophily based on few attributes can impede structural balance.
\newblock {\em Phys. Rev. Lett.}, 125:078302, 2020.

\bibitem{Hao2024}
Bingjie Hao and István~A. Kovács.
\newblock Proper network randomization is key to assessing social balance.
\newblock {\em Science Advances}, 10(18):eadj0104, 2024.

\bibitem{Gallo2024}
Anna Gallo, Diego Garlaschelli, Renaud Lambiotte, Fabio Saracco, and Tiziano
  Squartini.
\newblock Testing structural balance theories in heterogeneous signed networks.
\newblock {\em Communications Physics}, 7:154, 2024.

\bibitem{Kirkley2019}
Alec Kirkley, George~T. Cantwell, and M.~E.~J. Newman.
\newblock Balance in signed networks.
\newblock {\em Phys. Rev. E}, 99:012320, 2019.

\bibitem{Szell}
Michael Szell, Renaud Lambiotte, and Stefan Thurner.
\newblock Multirelational organization of large-scale social networks in an
  online world.
\newblock {\em Proceedings of the National Academy of Sciences},
  107(31):13636--13641, 2010.

\bibitem{Leskovec}
Jure Leskovec, Daniel Huttenlocher, and Jon Kleinberg.
\newblock Predicting positive and negative links in online social networks.
\newblock In {\em Proceedings of the 19th International Conference on World
  Wide Web}, pages 641--650, New York, NY, USA, 2010. Association for Computing
  Machinery.

\bibitem{antal2005social}
T.~Antal, P.~L. Krapivsky, and S.~Redner.
\newblock Dynamics of social balance on networks.
\newblock {\em Phys. Rev. E}, 72:036121, 2005.

\bibitem{ANTAL2006}
T.~Antal, P.L. Krapivsky, and S.~Redner.
\newblock Social balance on networks: The dynamics of friendship and enmity.
\newblock {\em Physica D: Nonlinear Phenomena}, 224(1):130--136, 2006.
\newblock Dynamics on Complex Networks and Applications.

\bibitem{marvel2009}
Seth~A. Marvel, Steven~H. Strogatz, and Jon~M. Kleinberg.
\newblock Energy landscape of social balance.
\newblock {\em Phys. Rev. Lett.}, 103:198701, 2009.

\bibitem{marvel2011continuous}
Seth~A Marvel, Jon Kleinberg, Robert~D Kleinberg, and Steven~H Strogatz.
\newblock Continuous-time model of structural balance.
\newblock {\em Proceedings of the National Academy of Sciences},
  108(5):1771--1776, 2011.

\bibitem{Belaza}
Andres~M. Belaza, Kevin Hoefman, Jan Ryckebusch, Aaron Bramson, Milan van~den
  Heuvel, and Koen Schoors.
\newblock Statistical physics of balance theory.
\newblock {\em PLOS ONE}, 12(8):1--19, 2017.

\bibitem{Hirotaka}
Hirotaka Goto, Masashi Shiraishi, and Hiraku Nishimori.
\newblock Onset of intragroup conflict in a generalized model of social
  balance.
\newblock {\em Phys. Rev. Lett.}, 133:127402, 2024.

\bibitem{Burda2022}
Zdzislaw Burda, Malgorzata~J. Krawczyk, and Krzysztof Ku\l{}akowski.
\newblock Perfect cycles in the synchronous heider dynamics in complete
  network.
\newblock {\em Phys. Rev. E}, 105:054312, 2022.

\bibitem{Gorski2023}
Piotr~J. Górski, Curtis Atkisson, and Janusz~A. Hołyst.
\newblock A general model for how attributes can reduce polarization in social
  groups.
\newblock {\em Network Science}, 11(4):536–559, 2023.

\bibitem{Oishi}
Koji Oishi and Kentaro Sakuwa.
\newblock Structural balance of alliance and rivalry networks in international
  relations.
\newblock {\em Artificial Life and Robotics}, 2024.

\bibitem{Dekker}
David Dekker, David Krackhardt, Patrick Doreian, and Pavel~N. Krivitsky.
\newblock Balance correlations, agentic zeros, and networks: The structure of
  192 years of war and peace.
\newblock {\em PLOS ONE}, 19(12):1--32, 12 2024.

\bibitem{Malarz}
Krzysztof Malarz, Maciej Wo\l{}oszyn, and Krzysztof Ku\l{}akowski.
\newblock Heider balance on archimedean lattices and cliques.
\newblock {\em Phys. Rev. E}, 111:014310, 2025.

\bibitem{Pham2020}
Minh~Tuan Pham, Imre Kondor, Rudolf Hanel, and Stefan Thurner.
\newblock The effect of social balance on social fragmentation.
\newblock {\em Journal of The Royal Society Interface}, 17(172):20200752, 2020.

\bibitem{Vendeville}
Antoine Vendeville and Fernando Diaz-Diaz.
\newblock Modeling echo chamber effects in signed networks.
\newblock {\em Phys. Rev. E}, 111:024302, 2025.

\bibitem{Pham2022}
Tuan~Minh Pham, Jan Korbel, Rudolf Hanel, and Stefan Thurner.
\newblock Empirical social triad statistics can be explained with dyadic
  homophylic interactions.
\newblock {\em Proceedings of the National Academy of Sciences},
  119(6):e2121103119, 2022.

\bibitem{Galesic2025}
Mirta Galesic, Henrik Olsson, Tuan Pham, Johannes Sorger, and Stefan Thurner.
\newblock Experimental evidence confirms that triadic social balance can be
  achieved through dyadic interactions.
\newblock {\em npj Complexity}, 2(1):1, 2025.

\bibitem{Kabashima_1998}
Y.~Kabashima and D.~Saad.
\newblock Belief propagation vs. {TAP} for decoding corrupted messages.
\newblock {\em Europhysics Letters}, 44(5):668, 1998.

\bibitem{yedidia2005constructing}
Jonathan~S Yedidia, William~T Freeman, and Yair Weiss.
\newblock Constructing free-energy approximations and generalized belief
  propagation algorithms.
\newblock {\em Information Theory, IEEE Transactions on}, 51(7):2282--2312,
  2005.

\bibitem{pearl1988probabilistic}
Judea Pearl.
\newblock {\em Probabilistic reasoning in intelligent systems: networks of
  plausible inference}.
\newblock Morgan Kaufmann, 1988.

\bibitem{kschischang2001factor}
Frank~R Kschischang, Brendan~J Frey, and Hans-Andrea Loeliger.
\newblock Factor graphs and the sum-product algorithm.
\newblock {\em IEEE Transactions on Information Theory}, 47(2):498--519, 2001.

\bibitem{Yedidia2000}
Jonathan~S Yedidia, William Freeman, and Yair Weiss.
\newblock Generalized belief propagation.
\newblock In T.~Leen, T.~Dietterich, and V.~Tresp, editors, {\em Advances in
  Neural Information Processing Systems}, volume~13, pages 689--695. MIT Press,
  2000.

\bibitem{yedidia2003understanding}
Jonathan~S Yedidia, William~T Freeman, and Yair Weiss.
\newblock Understanding belief propagation and its generalizations.
\newblock In {\em Exploring Artificial Intelligence in the New Millennium},
  pages 239--269. Morgan Kaufmann, 2003.

\bibitem{Rizzo2010}
Tommaso Rizzo, Alejandro Lage-Castellanos, Roberto Mulet, and Federico
  Ricci-Tersenghi.
\newblock Replica cluster variational method.
\newblock {\em Journal of Statistical Physics}, 139(3):375--416, 2010.

\bibitem{Metz2011}
F.~L. Metz, I.~Neri, and D.~Boll\'e.
\newblock Spectra of sparse regular graphs with loops.
\newblock {\em Phys. Rev. E}, 84:055101, 2011.

\bibitem{bolle2013spectra}
D{\'e}sir{\'e} Boll{\'e}, Fernando~Lucas Metz, and Izaak Neri.
\newblock On the spectra of large sparse graphs with cycles.
\newblock In H.~Holden et~al, editor, {\em Spectral {A}nalysis, {D}ifferential
  {E}quations and {M}athematical {P}hysics: {A} {F}estschrift in {H}onor of
  {F}ritz {G}esztesy’s 60th {B}irthday}, pages 35--58. American Mathematical
  Society, Providence, 2013.

\bibitem{Kirkley}
Alec Kirkley, George~T. Cantwell, and M.~E.~J. Newman.
\newblock Belief propagation for networks with loops.
\newblock {\em Science Advances}, 7(17):eabf1211, 2021.

\bibitem{Cantwell2023}
George~T. Cantwell, Alec Kirkley, and Filippo Radicchi.
\newblock Heterogeneous message passing for heterogeneous networks.
\newblock {\em Phys. Rev. E}, 108:034310, 2023.

\bibitem{WangPRL}
Yijia Wang, Yuwen~Ebony Zhang, Feng Pan, and Pan Zhang.
\newblock Tensor network message passing.
\newblock {\em Phys. Rev. Lett.}, 132:117401, 2024.

\bibitem{Guzman}
Grover E.~C. Guzman, Peter~F. Stadler, and Andre Fujita.
\newblock Cavity approach for the approximation of spectral density of graphs
  with heterogeneous structures.
\newblock {\em Phys. Rev. E}, 109:034303, 2024.

\bibitem{PhamPRE2024}
Tuan~Minh Pham, Thomas Peron, and Fernando~L. Metz.
\newblock Effects of clustering heterogeneity on the spectral density of sparse
  networks.
\newblock {\em Phys. Rev. E}, 110:054307, 2024.

\bibitem{Deffuant}
Guillaume Deffuant, David Neau, Frederic Amblard, and Gérard Weisbuch.
\newblock Mixing beliefs among interacting agents.
\newblock {\em Advances in Complex Systems}, 3(01n04):87--98, 2000.

\bibitem{Metropolis1953}
Nicholas Metropolis, Arianna~W. Rosenbluth, Marshall~N. Rosenbluth, Augusta~H.
  Teller, and Edward Teller.
\newblock Equation of state calculations by fast computing machines.
\newblock {\em The Journal of Chemical Physics}, 21(6):1087--1092, 06 1953.

\bibitem{Bahr1998}
David~B. Bahr and Eve Passerini.
\newblock Statistical mechanics of opinion formation and collective behavior:
  Micro‐sociology.
\newblock {\em The Journal of Mathematical Sociology}, 23(1):1--27, 1998.

\bibitem{MezardMontanariBook2009}
Marc Mézard and Andrea Montanari.
\newblock {\em Information, Physics, and Computation}.
\newblock Oxford University Press, 2009.

\bibitem{MezardParisi2001}
Marc M\'ezard and Giorgio Parisi.
\newblock The {B}ethe lattice spin glass revisited.
\newblock {\em The European Physical Journal B}, 20(2):217--233, 2001.

\bibitem{mozeika2009dynamical}
Alexander Mozeika and ACC Coolen.
\newblock Dynamical replica analysis of processes on finitely connected random
  graphs: Ii. dynamics in the griffiths phase of the diluted ising ferromagnet.
\newblock {\em Journal of Physics A: Mathematical and Theoretical},
  42(19):195006, 2009.

\bibitem{jan2025}
Jan Korbel, Shlomo Havlin, and Stefan Thurner.
\newblock Microscopic origin of abrupt mixed-order phase transitions.
\newblock {\em Nature Communications}, 16:2628, 2025.

\bibitem{Gorski2025}
Piotr~J Górski, Adam Sulik, Georges Andres, Giacomo Vaccario, and Janusz~A
  Hołyst.
\newblock Coexistence of balance and hierarchies: An ego perspective to explain
  empirical networks.
\newblock {\em PNAS Nexus}, 4(5):pgaf130, 04 2025.

\bibitem{Noudehi}
M.~Ghanbarzadeh Noudehi, A.~Kargaran, N.~Azimi-Tafreshi, and G.~R. Jafari.
\newblock Second- to first-order phase transition: Coevolutionary versus
  structural balance.
\newblock {\em Phys. Rev. E}, 106:044303, 2022.

\bibitem{Lage-Castellanos2008}
A.~Lage-Castellanos and R.~Mulet.
\newblock Zero temperature solutions of the edwards-anderson model in random
  husimi lattices.
\newblock {\em The European Physical Journal B}, 65(1):117--129, 2008.

\bibitem{Pedraza}
Luc\'{\i}a Pedraza, Sebasti\'an Pinto, Juan~Pablo Pinasco, and Pablo
  Balenzuela.
\newblock Analytical approach to the axelrod model based on similarity vectors.
\newblock {\em Phys. Rev. E}, 103:012307, 2021.

\bibitem{Montanari_2008}
Andrea Montanari, Federico Ricci-Tersenghi, and Guilhem Semerjian.
\newblock Clusters of solutions and replica symmetry breaking in random
  k-satisfiability.
\newblock {\em Journal of Statistical Mechanics: Theory and Experiment},
  2008(04):P04004, 2008.

\bibitem{Ricci-Tersenghi_2009}
Federico Ricci-Tersenghi and Guilhem Semerjian.
\newblock On the cavity method for decimated random constraint satisfaction
  problems and the analysis of belief propagation guided decimation algorithms.
\newblock {\em Journal of Statistical Mechanics: Theory and Experiment},
  2009(09):P09001, 2009.

\bibitem{zdeborova2010generalization}
Lenka Zdeborov{\'a} and Florent Krzakala.
\newblock Generalization of the cavity method for adiabatic evolution of gibbs
  states.
\newblock {\em Physical Review B—Condensed Matter and Materials Physics},
  81(22):224205, 2010.

\end{thebibliography}

\appendix

\onecolumngrid
\section{A. Derivation of Belief Propagation Equations}

To write the BP equations that characterize the system, we start from the full Hamiltonian. 
\begin{equation}
  H = \sum_i H^{(i)} = - \frac{\alpha}{g}\sum_{(i,j):J_{i,j}=1} {\bf s}_i \cdot {\bf s}_j + \frac{1-\alpha}{g}\sum_{(i,j):J_{i,j}=-1} {\bf s}_i \cdot {\bf s}_j
\end{equation}

Although this way of writing the Hamiltonian makes the physical interpretation easier, it will be useful here to recast everything as a more traditional sum over all the interacting pairs in the system:
\begin{equation}
  H = - \frac{\alpha}{2g}\sum_{<ij>} (1 + J_{i j}) ({\bf s}_i \cdot {\bf s}_j) + \frac{1-\alpha}{2g}\sum_{<ij>} (1-J_{ij})({\bf s}_i \cdot {\bf s}_j)
\end{equation}
where $J_{ij}$, as said in the main text, is the sign of the product ${\bf s}_i \cdot {\bf s}_j$. 

Both forms are equivalent, and we can use the explicit form of $J_{ij}$ and reorganize the terms to get:

\begin{equation}
  H = \frac{1-2\alpha}{2g}\sum_{<ij>} ({\bf s}_i \cdot {\bf s}_j) - \frac{1}{2g}\sum_{<ij>} |{\bf s}_i \cdot {\bf s}_j|
\end{equation}
where $|{\bf s}_i \cdot {\bf s}_j| = J_{ij}({\bf s}_i \cdot {\bf s}_j) = \text{sgn}({\bf s}_i \cdot {\bf s}_j) \, ({\bf s}_i \cdot {\bf s}_j)$ is the absolute value of the dot product.

Thus, the partition function $Z$ is:

\begin{eqnarray}
  Z &=& \sum_{\underline{\bf s} } e^{-\beta H(\underline{\bf s})} = \sum_{\underline{\bf s}} \prod_{<ij>} \exp \Big\{\frac{\beta}{2g} \Big[ (2 \alpha - 1) ({\bf s}_i \cdot {\bf s}_j) + \big|{\bf s}_i \cdot {\bf s}_j \big| \Big] \Big\} \nonumber \\
  Z &=& \sum_{\underline{\bf s}} \prod_{<ij>} \exp \Big\{\frac{\beta}{2g} ({\bf s}_i \cdot {\bf s}_j) \big[ (2 \alpha - 1) + \text{sgn}({\bf s}_i \cdot {\bf s}_j ) \big] \Big\}
\end{eqnarray}

The Boltzmann-Gibbs distribution of the different configurations is:

\begin{equation}
  P(\underline{\bf s}) = \frac{1}{Z} \prod_{<ij>} \exp \Big\{\frac{\beta}{2g} ({\bf s}_i \cdot {\bf s}_j) \big[ (2 \alpha - 1) + \text{sgn}({\bf s}_i \cdot {\bf s}_j ) \big] \Big\} = \frac{1}{Z} \prod_{<ij>} \varphi_{ij} ({\bf s}_i, {\bf s}_j)
\end{equation}
Once we have identified the shape of the pairwise factor $\varphi_{ij} ({\bf s}_i, {\bf s}_j)$, it is straightforward to obtain the BP equations by following the standard procedure \cite{yedidia2005constructing, Montanari_2008, Ricci-Tersenghi_2009, zdeborova2010generalization}:

   \begin{eqnarray}
  m_{j \rightarrow i}({\bf s}_j) \propto \prod_{k \in \partial j \setminus i} \sum_{{\bf s}_k} m_{k\rightarrow j}({\bf s}_k) \; \exp \Big\{\frac{\beta}{2g} ({\bf s}_k \cdot {\bf s}_j) \big[ 2 \alpha - 1 + {\rm sgn} ({\bf s}_k \cdot {\bf s}_j) \big] \Big\} \label{eq:BP_rule_app}
\end{eqnarray}

\section{B. Analytic prediction for some critical lines}

When the messages are initially set as random, we expect BP to stay with zero magnetization unless that disordered configuration is not a fixed point of the algorithm. To locate this transition, we study the stability of the disordered solution under small perturbations. This can be done in all the ensembles where the graphs are locally tree-like. 

Let us summarize the results and give the details of the computation later. The disordered solution stops being stable at the line defined as follows:
\begin{equation}
 \frac{Q(g, \beta, \alpha)}{R(g, \beta, \alpha)} = \frac{g}{2} \frac{1 + \langle \gamma \rangle}{\langle \gamma \rangle} \label{eq:critical_line}
\end{equation}
where $\gamma = c - 1$ is the reduced connectivity of a node, which is distributed according to $\mathbb{P}(\gamma)$, i.e. $\langle \gamma \rangle = \sum_{\gamma = 0}^{\infty} \gamma \, \mathbb{P}(\gamma)$. The shape of the functions $Q$ and $R$ depends on the parity of $g$. For even $g=2m$ we have:
\begin{eqnarray}
 R(2 m , \beta, \alpha) &=& e^{-\beta(\alpha-1)} A_{2m} \Big[e^{\beta(\alpha - 1) / m} \Big] + e^{\beta \alpha} A_{2m} \Big[e^{-\beta \alpha / m} \Big] + \binom{2 m}{m} \label{eq:R_2m} \\
 Q(2 m , \beta, \alpha) &=& e^{-\beta(\alpha-1)} e^{\beta(\alpha - 1) / m} A_{2m}' \Big[e^{\beta(\alpha - 1) / m} \Big] - e^{\beta \alpha} e^{-\beta \alpha / m} A_{2m}' \Big[e^{-\beta \alpha / m} \Big] + \nonumber \\
 & & + 2m e^{\beta \alpha} A_{2m} \Big[e^{-\beta \alpha / m} \Big] + m \, \binom{2 m}{m} \label{eq:Q_2m}
\end{eqnarray}
where, in terms of  the ordinary hypergeometric function ${}_2 F_1$, we have
\begin{eqnarray}
 A_{2m}(z) &=& (1 + z)^{2m} - \frac{m + 1}{m} \binom{2 m}{m - 1} z^{m} (1+z)^{2m} \,{}_2 F_1(2m + 1, m; m+1; -z) \label{eq:A2m_final} \\
 A_{2m}'(z) &=& \frac{2m}{1+z} A_{2m}(z) - (m + 1) \binom{2 m}{m - 1} \frac{z^{m - 1}}{1 + z} \label{eq:A2m_der}
\end{eqnarray}
On the other hand, for odd $g=2m + 1$ we obtain:

\begin{eqnarray}
 R(2 m + 1, \beta, \alpha) &=& e^{-\beta(\alpha-1)} A_{2m + 1} \Big[e^{2\beta(\alpha - 1) / (2 m + 1)} \Big] + e^{\beta \alpha} A_{2m + 1} \Big[e^{-2 \beta \alpha / (2m + 1)} \Big] \label{eq:R_2m1} \\
 Q(2 m + 1, \beta, \alpha) &=& e^{-\beta(\alpha-1)} e^{2\beta(\alpha - 1) / (2 m + 1)} A_{2m+1}' \Big[e^{2\beta(\alpha - 1) / (2 m + 1)} \Big] - \nonumber \\
 & & \!\!\!\!\!\!\!\!\!\!\!\!\!\!\!\!\!\!\!\!\!\!\!\!\!\!\!\!\!\!\!\! - e^{\beta \alpha} e^{-2 \beta \alpha / (2m + 1)} A_{2m}' \Big[e^{-2 \beta \alpha / (2m + 1)} \Big] +(2m + 1) e^{\beta \alpha} A_{2m+1} \Big[e^{-2 \beta \alpha / (2m + 1)} \Big] \label{eq:Q_2m1}
\end{eqnarray}
where

\begin{eqnarray}
 A_{2m+1}(z) &=& (1 + z)^{2m+1} - \binom{2 m+1}{m} z^{m+1} (1+z)^{2m+1} \,{}_2 F_1(2m + 2, m+1; m+2; -z) \label{eq:A_2m1} \\
 A_{2m+1}'(z) &=& \frac{2m+1}{1+z} A_{2m+1}(z) - (m + 1) \binom{2 m+1}{m} \frac{z^{m}}{1 + z} \label{eq:A2m1_der}
\end{eqnarray}

\subsection{B.1. Derivation}
Now let us proceed with the details of the computations needed to get our result in Eq. \eqref{eq:critical_line}. First, we rewrite the Eq. (\ref{eq:BP_rule_app}) considering explicitly the normalization factor $\mathcal{Z}_{j \to i}$:

 \begin{eqnarray}
  m_{j \rightarrow i}({\bf s}_j) & = &\frac{1}{\mathcal{Z}_{j \to i}} \prod_{k \in \partial j \setminus i} \sum_{{\bf s}_k} m_{k\rightarrow j}({\bf s}_k) \; \exp \Big\{\frac{\beta}{2g} ({\bf s}_k \cdot {\bf s}_j) \big[ 2 \alpha - 1 + \text{sgn} ({\bf s}_k \cdot {\bf s}_j) \big] \Big\} \label{eq:BP_rule_Z} \\
\mathcal{Z}_{j \to i} &= & \sum_{{\bf s_j}} \prod_{k \in \partial j \setminus i} \sum_{{\bf s}_k} m_{k\rightarrow j}({\bf s}_k) \; \exp \Big\{\frac{\beta}{2g} ({\bf s}_k \cdot {\bf s}_j) \big[ 2 \alpha - 1 + \text{sgn} ({\bf s}_k \cdot {\bf s}_j) \big] \Big\} \label{eq:Z_j_i}
\end{eqnarray}

In the region where there is no preferred orientation, the structure of the BP solutions corresponds to:

\begin{equation}
 m_{j \rightarrow i}({\bf s}_j) = \Big( \frac{1}{2} \Big)^{g} \:\:\:\:\:\:\:\:\:\:\:\:\:\:\:\:\:\:\:\: \forall {\bf s_j} \label{eq:dis_sol}
\end{equation}

To find the place where $m_{j \rightarrow i}({\bf s}_j)$ stops being a solution of the equations, we make a small perturbation to each term of the product in Eq. \ref{eq:dis_sol}:

\begin{equation}
 m_{j \rightarrow i}({\bf s}_j) = \prod_{l=1}^{g} \Big( \frac{1 + s_j^{l} \mu_{j \to i}^{l}}{2} \Big) \label{eq:ansatz_m_1}
\end{equation}

Here, $s_j^{l}$ is the $l^{\text{th}}$ component of ${\bf s}_j$ and $\mu_{j \to i}^{l}$ is the average value of that $l^{\text{th}}$ component. Notice that the distribution in Eq. \ref{eq:ansatz_m_1} is still normalized, and we assume that it remains factorized in the different components. In the completely disordered solution, we have $\mu_{j \to i}^{l}=0$. Thus, this will be the small parameter in the perturbation we are making. 

There is no reason to believe that one direction should be preferred. Thus, we expect a sort of spontaneous symmetry breaking in which the system randomly chooses to orient itself into one of the directions. To fix ideas, let us say that the preferred direction will have $s_j^{l}=1$ for all $l=1, \ldots, g$. In that case $\mu_{j \to i}$ is the same for all components and Eq.(\ref{eq:ansatz_m_1}) transforms into:

\begin{equation}
 m_{j \rightarrow i}({\bf s}_j) = \Big( \frac{1 + \mu_{j \to i}}{2} \Big)^{n_{j \to i}} \, \Big( \frac{1 - \mu_{j \to i}}{2} \Big)^{g - n_{j \to i}} \label{eq:ansatz_m_n}
\end{equation}
where $n_{j \to i} = \sum_{l=1}^{g} \delta(s_j^{l}, 1)$ is the number of components of ${\bf s}_j$ that point upward.

By inserting (\ref{eq:ansatz_m_n}) into (\ref{eq:BP_rule_Z}) and averaging over the graph ensemble, one gets:

\begin{equation}
 m_{\text{new}}({\bf s}^{\ast}) = \sum_{\gamma = 0}^{\infty} P(\gamma) \, \frac{N_{\gamma}(g, \beta, \alpha)}{Z_{\gamma}(g, \beta, \alpha)} \label{eq:av_gamma}
\end{equation}
 The numerator $N_{\gamma}(g, \beta, \alpha)$ and the denominator $Z_{\gamma}(g, \beta, \alpha)$ are given by:

\begin{eqnarray}
 N_{\gamma}(g, \beta, \alpha) &=& \prod_{k = 1}^{\gamma} \sum_{n_k=0}^{g} \binom{g}{n_k} \Big( \frac{1 + \mu}{2} \Big)^{n_{k}} \, \Big( \frac{1 - \mu}{2} \Big)^{g - n_{k}} \times \nonumber \\
 & & \times \exp \Big\{\frac{\beta}{2g} (2 n_k - g) \big[ 2 \alpha - 1 + \text{sgn} (2 n_k - g) \big] \Big\} \label{eq:num_exp} \\
 Z_{\gamma}(g, \beta, \alpha) &=& \sum_{n_j=0}^{g} \binom{g}{n_j} \prod_{k = 1}^{\gamma} \sum_{n_k^{+}=0}^{n_j} \binom{n_j}{n_k^{+}} \sum_{n_k^{-}=0}^{g - n_j} \binom{g - n_j}{n_k^{-}} \Big( \frac{1 + \mu}{2} \Big)^{n_k^{+} + g - n_j - n_k^{-}} \, \Big( \frac{1 - \mu}{2} \Big)^{n_j + n_k^{-} - n_k^{+}} \times \nonumber \\
 & & \times \exp \Big\{\frac{\beta}{2g} (2 n_k^{+} + 2 n_k^{-} - g) \big[ 2 \alpha - 1 + \text{sgn} (2 n_k^{+} + 2 n_k^{-} - g) \big] \Big\} \label{eq:den_exp}
\end{eqnarray}

In Eq. (\ref{eq:av_gamma}), we have chosen $n_{j \to i} = g$ and, therefore, all the components of the vector ${\bf s}^{\ast}$ are equal to one. The message $m_{\text{new}} ({\bf s}^{\ast})$ is the one obtained after applying the update rule to the perturbed messages. The reader should recall that we are studying the stability of the fixed point $\hat{m}({\bf s})=(1/2)^{g}$ for all ${\bf s}$. We need to understand whether, after applying the small perturbation, $m_{\text{new}}({\bf s}^{\ast})$ goes back in the direction of the fixed point or, on the contrary, goes away from it.

Thus, the idea is to expand $m_{\text{new}}({\bf s}^{\ast})$ around $\hat{m}({\bf s})=(1/2)^{g}$ in powers of $\mu$. The fixed point is unstable if

\begin{equation}
  \Big| \Big(\frac{1+\mu}{2}\Big)^{g}- \Big(\frac{1}{2} \Big)^{g} \Big| \leq \Big| m_{\text{new}}({\bf s}^{\ast}) - \Big(\frac{1}{2} \Big)^{g} \Big|
\end{equation}

Expanding the left-hand side of this inequality in powers of $\mu$, keeping only the first-order terms and taking the equality, one gets:

\begin{equation}
  \Big(\frac{1}{2} \Big)^{g} (1 + g \mu) = \Big| m_{\text{new}}({\bf s}^{\ast}) - \Big(\frac{1}{2} \Big)^{g} \Big| \label{eq:stab_line}
\end{equation}

We will estimate the stability line of the paramagnetic state by finding the solutions of Eq. \ref{eq:stab_line}. To help the reader follow the computation, let us interpret Eqs. \ref{eq:num_exp} and \ref{eq:den_exp}. Notice that we are giving an expression for a message between two sites, say $i$ and $j$. We are using all the incoming messages from all the other neighbors of $j$. The numerator (\ref{eq:num_exp}) is a product of sums, and each factor in the product is the contribution of one of the $\gamma=c-1$ neighbors. They all become equivalent because we are assuming all the messages to be equal.

The sum inside the product in Eq. \ref{eq:num_exp} goes over all the possible configurations of one of these neighbors. To fix ideas, let us say that the neighbor has an opinion ${\bf s}_k$. Since what is important is the scalar product between the opinion ${\bf s}_j$ and the opinion of the neighbor ${\bf s}_k$, we care about how many of the $g$ components of ${\bf s}_k$ are equal to the corresponding components of ${\bf s}_j$. That number is what we call $n_k$ in Eq. \ref{eq:num_exp}. The sum over all the configurations of ${\bf s}_k$ is substituted for a sum over all the values of $n_k$, with a combinatorial factor that counts the configurations with the same scalar product ${\bf s}_j \cdot {\bf s}_k = 2n_k - g$. In Eq. \ref{eq:num_exp} we also used Eq. \ref{eq:ansatz_m_n} for the messages, remembering that we set all the components of ${\bf s}_j$ to one.

The denominator in Eq. \ref{eq:den_exp} is just the normalization of the numerator (Eq. \ref{eq:num_exp}). We consider all the values of ${\bf s}_j$ by summing over $n_j$, which is the number of positive components of ${\bf s}_j$. Inside the outer sum, the structure is the same. The integer $n_k^{+}$ is the number of components of ${\bf s}_k$ whose value is '1' and that are also equal to the corresponding components of ${\bf s}_j$. We must have $n_k^{+} \leq n_j$. On the other hand, $n_k^{-}$ is the number of components of ${\bf s}_k$ whose value is '-1' and that are also equal to the corresponding components of ${\bf s}_j$. We have $n_k^{-}\leq g - n_j$. Therefore, the scalar product is ${\bf s}_j \cdot {\bf s}_k = 2n_k^{+}+2n_k^{-} - g$. The total number of components of ${\bf s}_k$ whose value is '1' is $n_k^{+} + g - n_j - n_k^{-}$, and the other $n_j + n_k^{-} - n_k^{+}$ are '-1'.

Now, we can expand both equations in powers of $\mu$. Let us take the numerator (Eq. \ref{eq:num_exp}) and keep only the terms up to first order:

\begin{eqnarray}
N_{\gamma}(g, \beta, \alpha) &=& 2^{-\gamma \, g} \prod_{k = 1}^{\gamma} \sum_{n_k=0}^{g} \binom{g}{n_k} \Big( 1 + (2n_k - g)\mu \Big) \times \nonumber \\
 & & \times \exp \Big\{\frac{\beta}{2g} (2 n_k - g) \big[ 2 \alpha - 1 + \text{sgn} (2 n_k - g) \big] \Big\} + o(\mu) \nonumber \\
 N_{\gamma}(g, \beta, \alpha) &=& 2^{-\gamma \, g} \prod_{k = 1}^{\gamma} \, \big[ \, (1-g\mu)R(g, \beta, \alpha) + 2\mu Q(g, \beta, \alpha) \, ] + o(\mu) \nonumber \\
 N_{\gamma}(g, \beta, \alpha) &=& 2^{-\gamma \, g} \big[ \, (1-\gamma g\mu)R(g, \beta, \alpha) + 2\mu Q(g, \beta, \alpha) \, \big]^{\gamma} + o(\mu) \nonumber \\
 N_{\gamma}(g, \beta, \alpha) &=& 2^{-\gamma \, g} \Big(R(g, \beta, \alpha) \Big)^{\gamma - 1} \big[ \, (1-\gamma\, g \, \mu)R(g, \beta, \alpha) + 2\mu Q(g, \beta, \alpha) \, \big] + o(\mu) \label{eq:num_exp_first_order}
\end{eqnarray}
where we defined the sums:

\begin{eqnarray}
R(g, \beta, \alpha) &=& \sum_{n_k=0}^{g} \binom{g}{n_k} \exp \Big\{\frac{\beta}{2g} (2 n_k - g) \big[ 2 \alpha - 1 + \text{sgn} (2 n_k - g) \big] \Big\} \label{eq:S1} \\
 Q(g, \beta, \alpha) &=& \sum_{n_k=0}^{g} \binom{g}{n_k} \,n_k \, \exp \Big\{\frac{\beta}{2g} (2 n_k - g) \big[ 2 \alpha - 1 + \text{sgn} (2 n_k - g) \big] \Big\} \label{eq:S2}
\end{eqnarray}

To compute $R(g, \beta, \alpha)$ and $Q(g, \beta, \alpha)$, we need to deal with the 'sign' function inside the exponential. The straightforward choice is to split the sum into two or possibly three parts: when the sign is negative, positive, or zero. 

When $g$ is even, the sign can be zero. Let us take $m=g/2$ and write:

\begin{eqnarray}
R(2m, \beta, \alpha) &=& \sum_{n_k=0}^{m - 1} \binom{2m}{n_k} \exp \Big\{\frac{\beta}{m} (n_k - m) (\alpha - 1) \Big\} + \binom{2m}{m} + \sum_{n_k=m+1}^{2m} \binom{2m}{n_k} \exp \Big\{\frac{\beta}{m} (n_k - m) \alpha \Big\} \nonumber \\
R(2m, \beta, \alpha) &=& e^{-\beta(\alpha-1)} \sum_{n_k=0}^{m - 1} \binom{2m}{n_k} \exp \Big\{\frac{\beta(\alpha -1)}{m} \, n_k \Big\} + \binom{2m}{m} + e^{\beta \, \alpha} \sum_{n_k=0}^{m - 1} \binom{2m}{n_k} \exp\Big\{-\frac{\beta \, \alpha}{m} \, n_k\Big\} \label{eq:S1_2m}
\end{eqnarray}

Then, we identify that

\begin{eqnarray}
R(2m, \beta, \alpha) &=& e^{-\beta(\alpha-1)} A_{2m} \Big[e^{\beta(\alpha - 1) / m} \Big] + e^{\beta \alpha} A_{2m} \Big[e^{-\beta \alpha / m} \Big] + \binom{2 m}{m} \label{eq:S1_2m_A}
\end{eqnarray}
where

\begin{eqnarray}
A_{2m}[z] &=& \sum_{n=0}^{m - 1} \binom{2m}{n} z^{n} \label{eq:A2m_def}
\end{eqnarray}

To compute this last sum, we take the derivative of Eq. (\ref{eq:A2m_def}) with respect to $z$:

\begin{eqnarray}
A_{2m}'[z] &=& \sum_{n=0}^{m - 1} n \binom{2m}{n} z^{n-1}= \sum_{n=0}^{m - 2} \frac{(2m)! \, z^{n}}{n!(2m-n-1)!} = \sum_{n=0}^{m - 1} (2m - n) \frac{(2m)! \, z^{n}}{n!(2m-n)!} - (m + 1) \frac{(2m)! \, z^{m-1}}{(m -1 )!(m+1)!} \nonumber \\
A_{2m}'[z] &=& 2m A_{2m}[z] - z A_{2m}'[z] - (m + 1) \binom{2m}{m-1} \, z^{m-1} \label{eq:diff_eq_A2m}
\end{eqnarray}
from where

\begin{eqnarray}
(1+z)^{2m} \frac{d}{dz}\Big[ (1+z)^{-2m} A_{2m}[z] \Big] &=& - (m + 1) \binom{2m}{m-1} \, \frac{z^{m-1}}{1+z} \nonumber \\ (1+z)^{2m} A_{2m}[z] &=& (1+z)^{2m} - (m + 1) \binom{2m}{m-1} \, (1+z)^{2m} \, \int_0^{x} dt \, \frac{t^{m-1}}{(1+t)^{2m+1}} \label{eq:desp_eq_A2m}
\end{eqnarray}
where we used that $A_{2m}[0]=1$.

Taking into account now the following integral expression of the ordinary hypergeometric function:

\begin{eqnarray}
  {}_2F_{1}(a,b;c;z)=\frac{1}{B(b,c-b)}\int_0^{1} x^{b-1} (1-x)^{-b-1}(1-zx)^{-a} dx \:\:\:\:\:\:\:\:\:\:\:\: Re[c] > Re[b] > 0
\end{eqnarray}
where $B(b,c-b)$ is Euler's Beta function, we get Eq. \ref{eq:A2m_final}.

\begin{eqnarray}
A_{2m}(z) &=& (1 + z)^{2m} - \frac{m + 1}{m} \binom{2 m}{m - 1} z^{m} (1+z)^{2m} \,{}_2 F_1(2m + 1, m; m+1; -z) \label{eq:A2m_final_2}  
\end{eqnarray}

Once we see how our computations are connected to the hypergeometric function, the rest is analogous. It is not difficult to see that for even $g$ the sum $Q(g, \beta, \alpha)$ is then related to the derivative of $A_{2m}[z]$ (see Eqs. \ref{eq:Q_2m} and \ref{eq:A2m_der}). When $g$ is odd, the procedure is the same with only small modifications that are reflected in the results presented in Eqs. \ref{eq:R_2m1}, \ref{eq:Q_2m1}, \ref{eq:A_2m1}, and \ref{eq:A2m1_der}.

When we insert everything back into the Eq. \ref{eq:av_gamma} and keep expanding in powers of $\mu$ to the first order, we get:

\begin{equation}
 1 + g \mu = 1 + \langle \gamma \rangle \Big(\frac{2 Q(g, \beta, \alpha)}{R(g, \beta, \alpha)} - g \Big) \, \mu + o(\mu) \label{eq:expansion}
\end{equation}
where $\langle \gamma \rangle = \sum_{\gamma = 0}^{\infty} \gamma \, \mathbb{P}(\gamma)$. 
Therefore, we arrive at Eq. \ref{eq:critical_line}:

\begin{equation}
 \frac{Q(g, \beta, \alpha)}{R(g, \beta, \alpha)} = \frac{g}{2} \frac{1 + \langle \gamma \rangle}{\langle \gamma \rangle} \label{eq:critical_line_2}
\end{equation}

\section{C. Capturing the transition with Belief Propagation}

 There are $2 \times 2^{g}$ equations like Eq. \ref{eq:BP_rule_app} for each edge in the interaction graph. We can solve them by iterating until the convergence of the messages $m_{i \to j}({\bf s}_j)$. If our graph is locally tree-like, we should expect a good agreement between the results of this Belief Propagation (BP) algorithm and the Monte Carlo simulations. Since we suspect the presence of a transition between two different states, one where the opinion vectors ${\bf s}_i$ do not have any preferential orientation, and other where they exhibit a spontaneous alignment towards a preferred direction, 
 we will compute the absolute value of the energy density $| e |$, where $e$ is also known as the level of social stress per pair of agents, the fraction of positive links $\rho_{+}$, and the magnetization $m$. Their definitions are:

 \begin{eqnarray}
 | e | &=& -\langle H \rangle = \frac{1}{L} \sum_{\langle ij \rangle} \sum_{{\bf s}_i} \sum_{{\bf s}_j} \, P({\bf s}_i, {\bf s}_j) \, \frac{{\bf s}_i \cdot {\bf s}_j}{2 G} \big[2 \alpha - 1 + \text{sign}({\bf s}_i \cdot {\bf s}_j) \big] \nonumber \\
 \rho_{+} &=& \frac{1}{L} \sum_{\langle ij \rangle} \sum_{{\bf s}_i} \sum_{{\bf s}_j} \, P({\bf s}_i, {\bf s}_j) \, \Theta\big( {\bf s}_i \cdot {\bf s}_j \big) \nonumber \\
 m &=& \Big|\Big| \frac{1}{N} \sum_{i} \sum_{{\bf s}_i} P({\bf s}_i) {\bf s}_i \Big| {\Big|}_{2}
 \end{eqnarray}
where $P({\bf s}_i)$ and $P({\bf s}_i, {\bf s}_j)$ can be computed using BP or Monte Carlo, $L$ is the number of edges in the graph and $N$ the number of vertices, $\Theta(x)$ is one only if $x > 0$ and gives zero otherwise, and $ | | \, {\bf v} \, | |_{2}$ is the Euclidean norm of the vector ${\bf v}$.

 It can be easily checked that the maximum value of $| e |$ depends on $\alpha$. For $\alpha = 1 / 2$ we have 

\begin{eqnarray}
 | e | = \frac{1}{L} \sum_{\langle ij \rangle} \frac{ \mathbb{E}_{ij} \big[ \, | {\bf s}_i \cdot {\bf s}_j | \, \big] }{2 G} \leq \frac{1}{2} \nonumber
 \end{eqnarray}
 where $\mathbb{E}_{ij} [ \,\cdot \,] \equiv \sum_{{\bf s}_i} \sum_{{\bf s}_j} \, P({\bf s}_i, {\bf s}_j) \, [ \,\cdot \,]$. For $\alpha = 1$ we have
 
 \begin{eqnarray}
 | e | = \frac{1}{L} \sum_{\langle ij \rangle} \Big( \frac{ \mathbb{E}_{ij} \big[ {\bf s}_i \cdot {\bf s}_j \big] }{2 G} + \frac{ \mathbb{E}_{ij} \big[ \, | {\bf s}_i \cdot {\bf s}_j | \, \big] }{2 G} \Big) \leq 1 \nonumber
 \end{eqnarray}

\begin{figure}[H]
\centering
 \subfloat[]{
	  \centering
	  \includegraphics[width=0.45\textwidth]{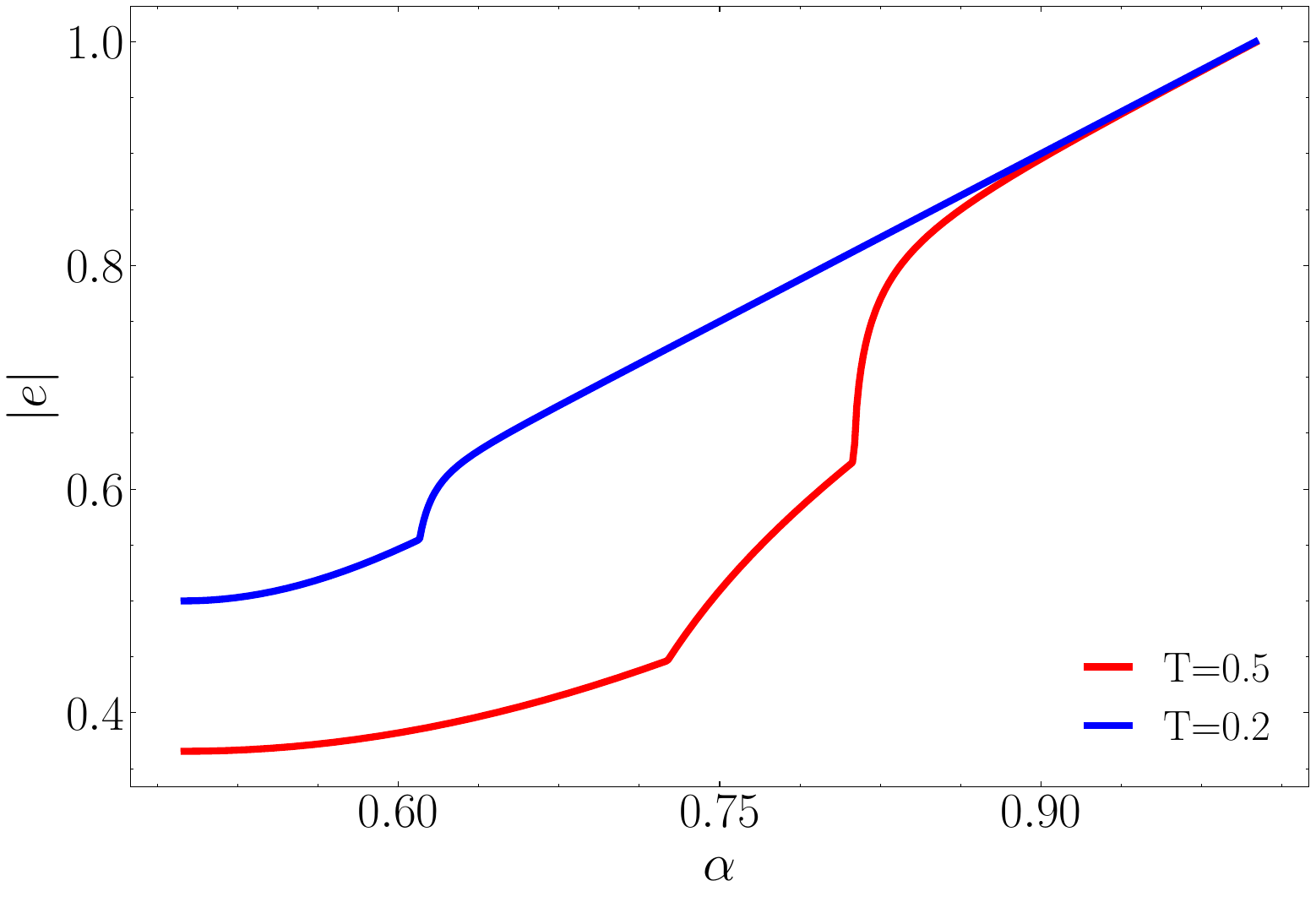}}
            \subfloat[]{
	  \centering
	  \includegraphics[width=0.45\textwidth]{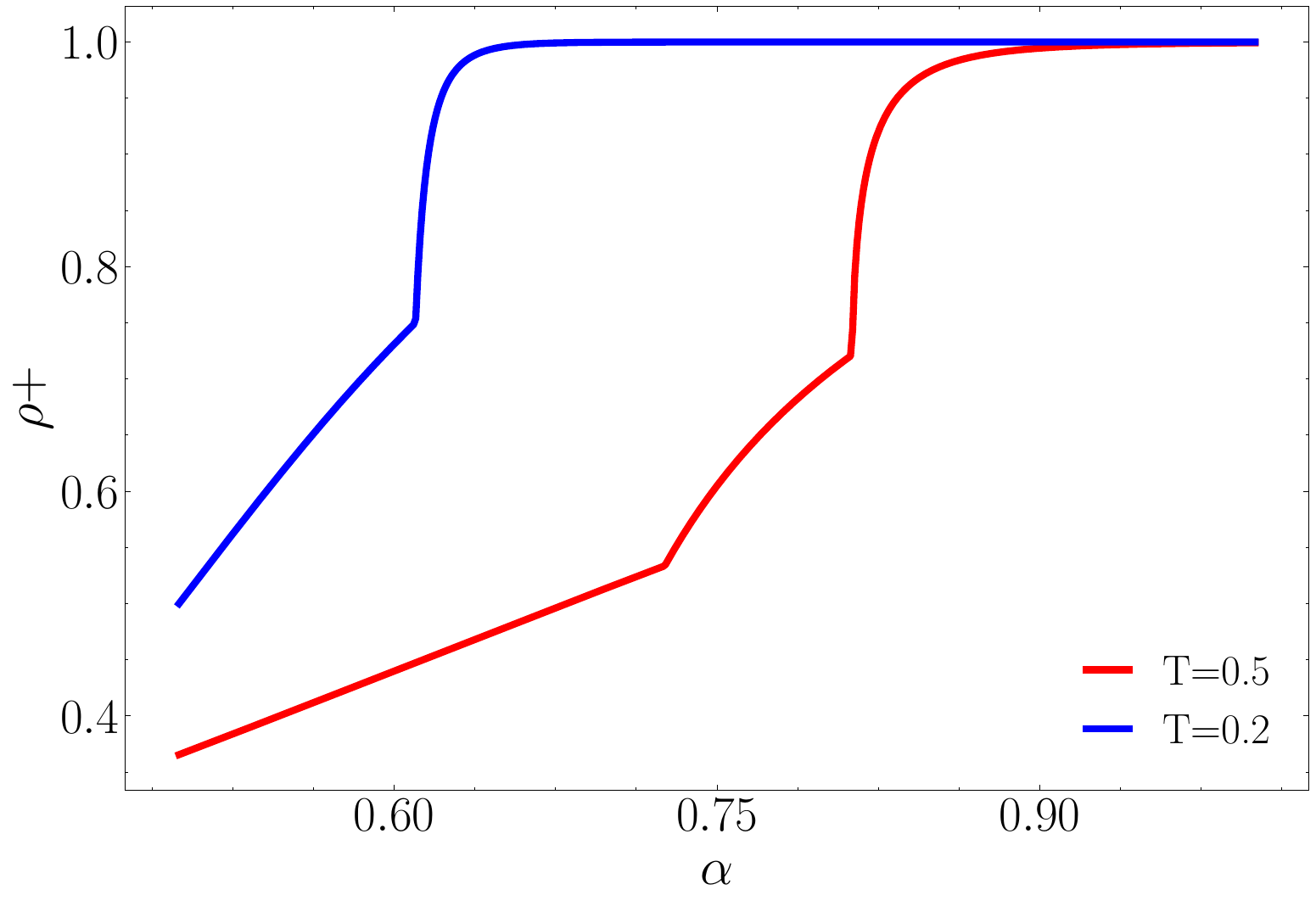}}
	  
\caption{Belief Propagation results for the absolute value of the energy density $|e|$ and the fraction of positive links $\rho_{+}$ in random regular graphs with connectivity $c=3$. The lines are the outcome of BP's convergence process for a given value of $\alpha$ and the temperature $T$. All the messages are initially aligned towards a given direction. The panels contain results $g=2$ and two temperatures, $T=1/5$ and $T=1/2$.}
\label{fig:BP_single_instance_e}
\end{figure}

 In Fig. \ref{fig:BP_single_instance_e} we show the values of $|e|$ and $\rho_+$ for different values of $\alpha$ around the transition. The calculations were performed for random regular graphs. In that case, if one chooses the same initial values for all the messages in a graph, the set of BP equations reduces to only one equivalent equation. Therefore, the numerics can be directly done using only one equation, and it is very fast.
 
\begin{figure}[H]
\centering
	  \includegraphics[width=0.45\textwidth]{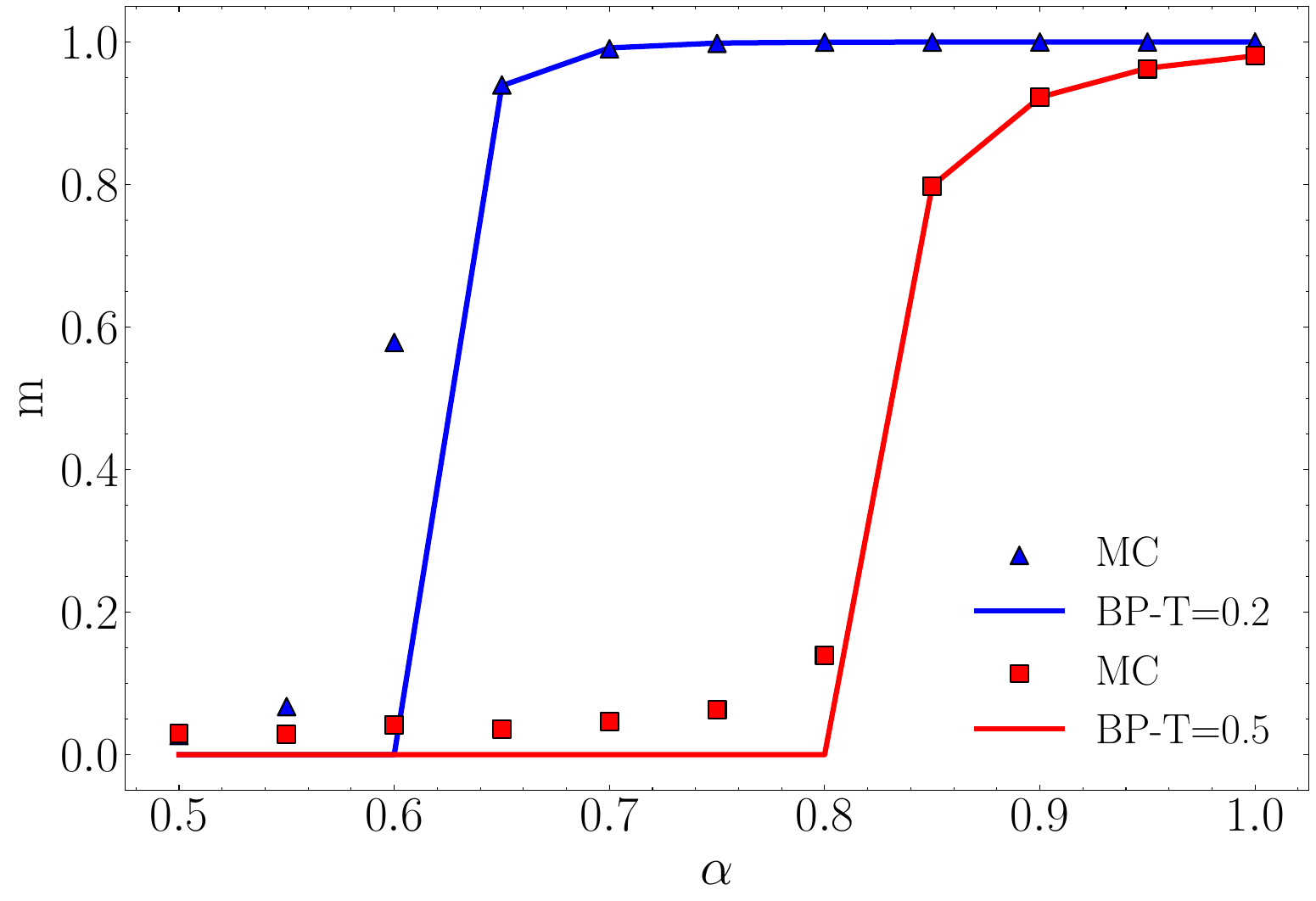}
\caption{Comparison between Monte Carlo's and BP's magnetization in random regular graphs with connectivity $c=3$. The number of individual opinions is $g=2$ in all cases. Each point is the result of averaging $s=50$ Monte Carlo simulations in systems of size $N=1000$. All simulations started with all the spins aligned towards a given direction ($m(t=0)=1$). The lines are the outcome of BP's convergence for some values of $\alpha$ and the temperature $T$. All the messages are set initially to favor the same direction chosen for Monte Carlo's initial conditions. The figure contains results for two temperatures, $T=1/5$ and $T=1/2$.}
\label{fig:BP_vs_MC_single_instance_M}
\end{figure}

Although the energy and fraction of positive links show some small changes in behavior around specific values of $\alpha$, it is not possible to see the extent of what is happening in the system. Therefore, we also computed the magnetization in Fig. \ref{fig:BP_vs_MC_single_instance_M}, where we find a good agreement between Monte Carlo simulations and the results of the BP equations. The figure also shows the existence of a sharp temperature-dependent change in the system's behavior around a specific value of $\alpha$: below a certain $\alpha$ there is no preferred orientation, and above that $\alpha$ the system aligns.

The sharp transitions observed in the magnetization coincide with discontinuities in the derivative with respect to $alpha$ of the energy and the fraction of positive links. Notice that at $T=0.2$ there is only one discontinuity in the derivative (blue curves in Fig. \ref{fig:BP_single_instance_e}). However, at $T=0.5$ both the energy and the fraction of positive links display two discontinuities in their derivatives with respect to $\alpha$. The one located at the larger $\alpha$ is explained by the order-disorder transition signaled by the magnetization in Fig. \ref{fig:BP_vs_MC_single_instance_M}. The other one, at a smaller $\alpha$, does not seem to be related to this transition, and further investigation is needed to understand its origin.

\section{D. Monte Carlo dynamics}

\begin{figure}[H]
\centering
 \subfloat[]{
	  \centering
	  \includegraphics[width=0.45\textwidth]{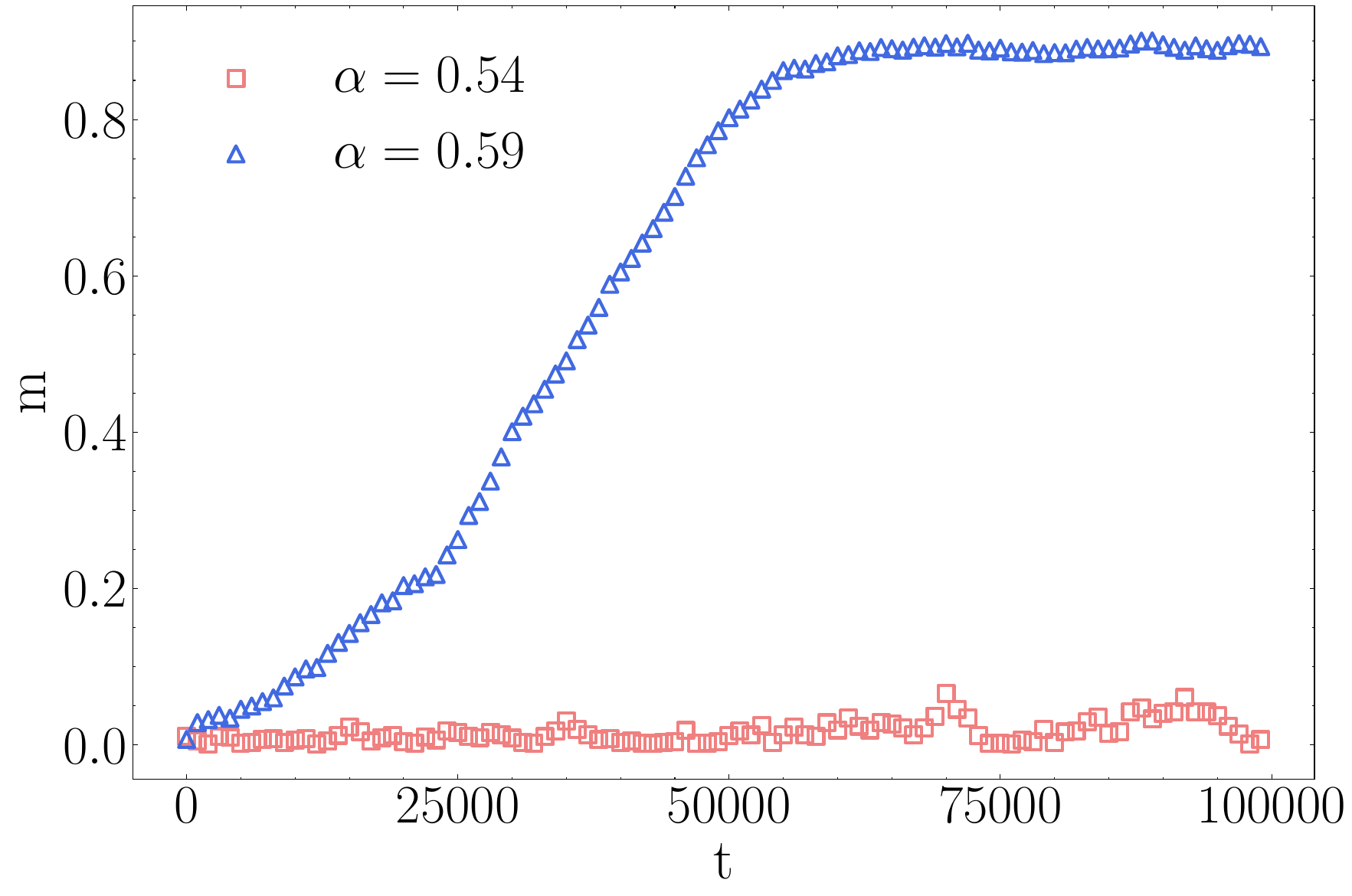}}
      \subfloat[]{
	  \centering
	  \includegraphics[width=0.45\textwidth]{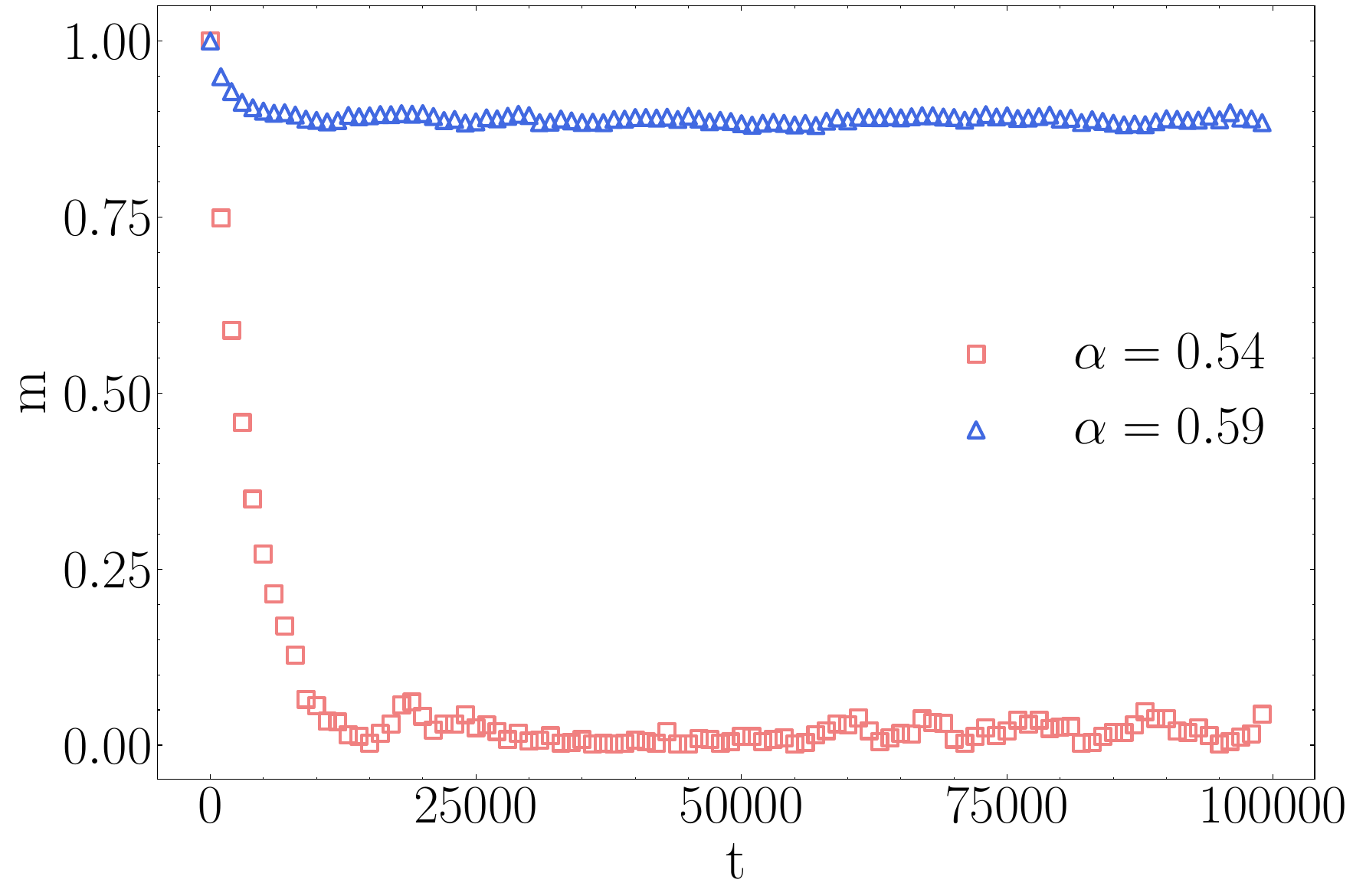}}
    
 \subfloat[]{
	  \centering
	  \includegraphics[width=0.45\textwidth]{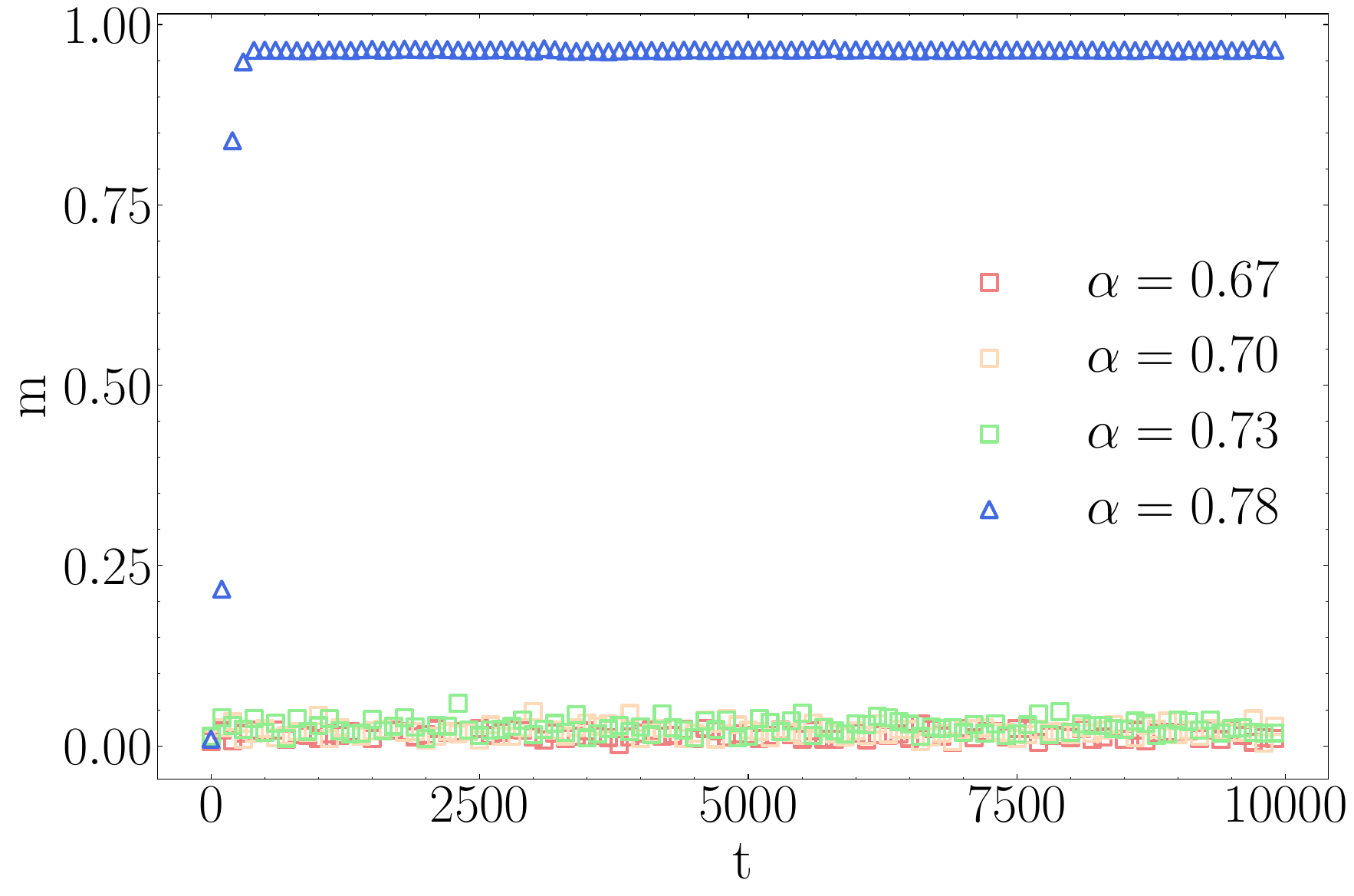}}
      \subfloat[]{
	  \centering
	  \includegraphics[width=0.45\textwidth]{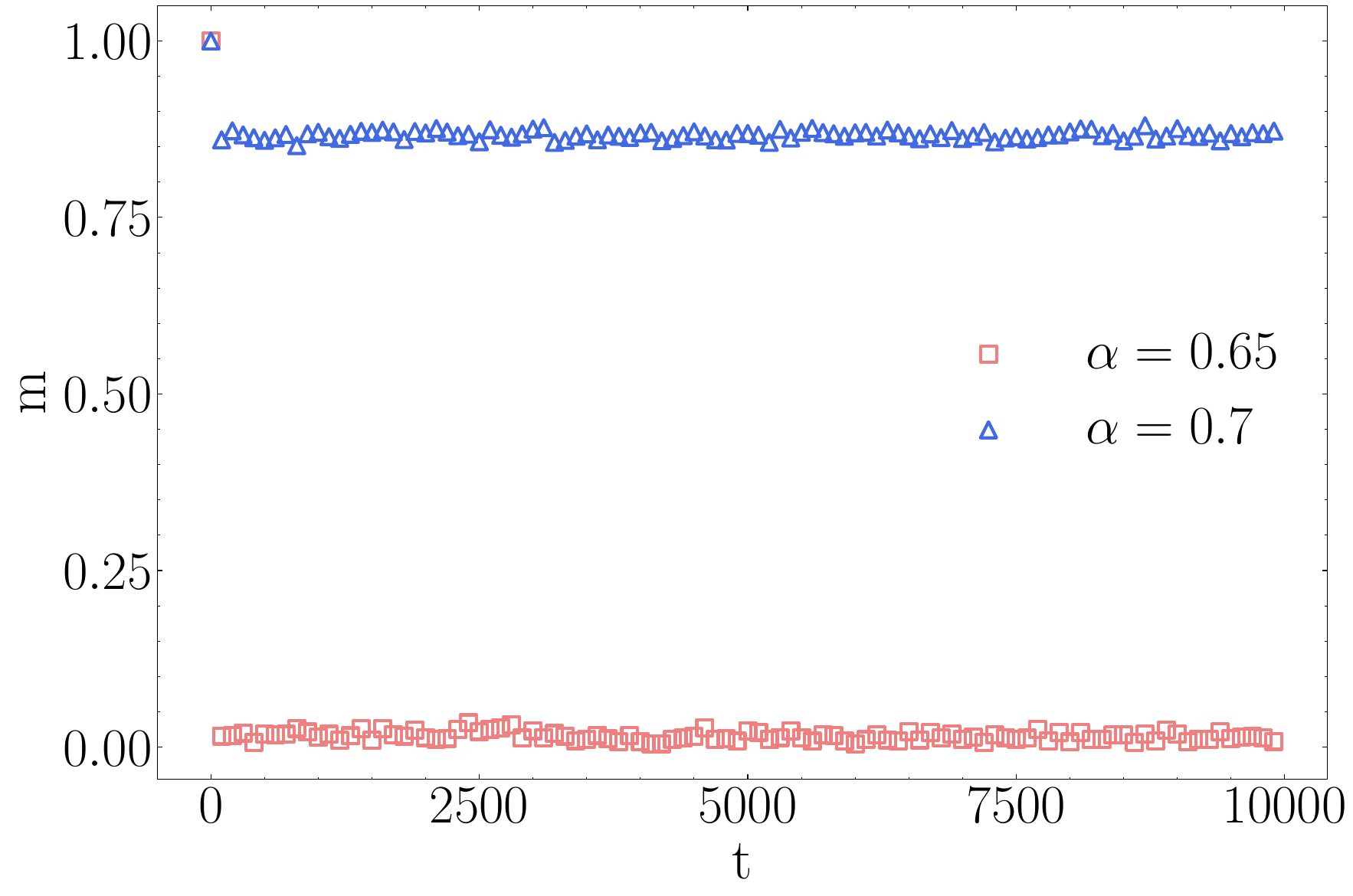}}
\caption{Monte Carlo dynamics in random regular graphs with connectivity $c=4$. The length of the opinions of each individual is $g=4$. The upper panels show the average of $5$ Monte Carlo runs at a low temperature ($T=0.19$). The bottom panels show the average of $20$ Monte Carlo runs at a higher temperature ($T=0.33$). The simulations in the left panels start with a random initial condition, while the ones in the right panels start completely aligned towards a given direction. System size is $N=10000$. The time unit is a Monte Carlo step, consisting of N individual Metropolis steps.}
\label{fig:MC_dyn}
\end{figure}

The results obtained with BP suggest the relevance of the initial conditions in the system's behavior. In particular, the location of the transition strongly depends on the initial conditions. We show in Fig. (\ref{fig:MC_dyn}) that this is also the case for Monte Carlo simulations. Here, the upper panels show Monte Carlo runs performed at a low temperature ($T=0.19$), while the lower panels are done for a high temperature ($T=0.33$). The left column presents results obtained from a random initial condition with small magnetization, and the right column shows those obtained from a completely aligned initial condition. 

For $T=0.19$, the steady states are independent of the initial conditions. On the other hand, for $T=0.33$, the steady states \emph{do} depend on the initial condition. For example, in the bottom-left panel, the system is not magnetized at any time for $\alpha=0.7$, while in the bottom-right the same system remains magnetized for a long time. Moreover, the blue points in the bottom-left panel obtained at a larger value $\alpha = 0.73$ are still close to $m=0$. 

\section{E. Behavior of the area of possible consensus when the number of discussed issues increases}

\begin{figure}[H]
    \centering
    \includegraphics[width=0.45\linewidth]{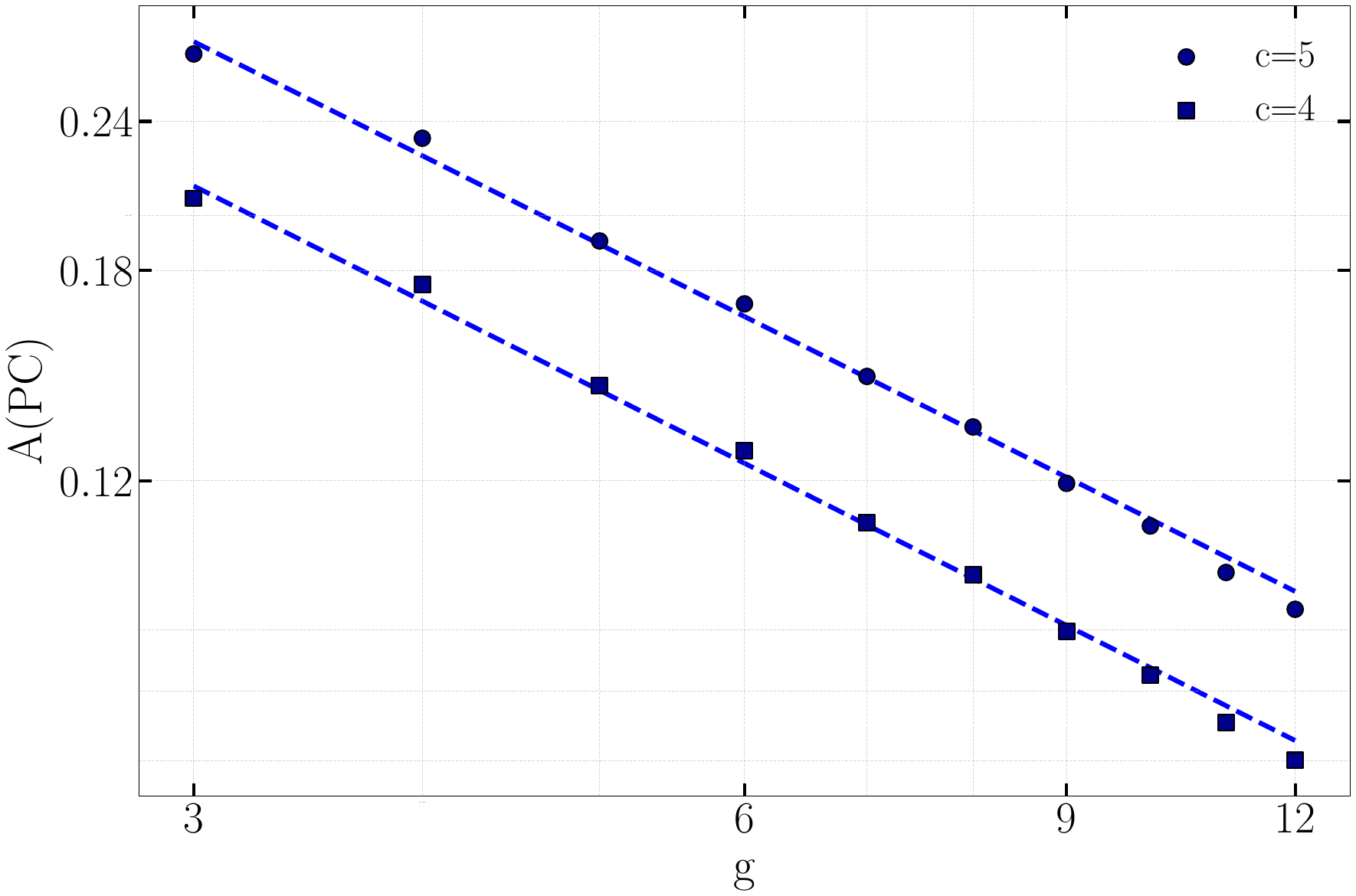}
    \caption{Area of possible consensus A(PC) \textit{vs.} the length $g$ of the opinion vectors on a logarithmic scale with fixed connectivity $c= 4$ (squares) and $c=5$ (circles). Blue dashed lines are fitted to the data points using a power-law $A(EC) = \Phi(c) \,g^{-\gamma(c)}$. The optimal parameters depend on the connectivity. They are: $\Phi(4)=0.494(15)$, $\Phi(5)=0.61(2)$, and $\gamma(4)=0.771(18)$, $\gamma(5)=0.764(18)$.}
    \label{fig:APC}
\end{figure}

In the main text, we identify two relevant regions of the model's phase diagram. We denoted by A(EC) the area where consensus is easy to find, regardless of the initial conditions. We showed that A(EC) decreases when the number of discussed issues ($g$) increases, following a power-law. On the other hand, A(PC) (area of possible consensus) is the entire area where consensus is possible. For completeness, in this section we include Fig. \ref{fig:APC}, illustrating the dependence of A(PC) on the value of $g$. The area where consensus is possible also decays with $g$, but this cannot be strictly described using a power-law. For small $g$, the points in Fig. \ref{fig:APC}, displayed in logarithmic scale, are close to a straight line representing the best power law fit to the points. For large $g$, instead, the points start to deviate from this power law.

Two things must be commented on here. First, the value of A(PC) is not as relevant as the value of A(EC). Only the latter represents the region where consensus can be reached from a random initial condition, \textit{i.e.}, A(EC) is more suitable as a quantitative measure of consensus feasibility. Second, the fact that A(PC) is not a power-law is responsible for the saturation of the gap size (see Fig. 2 in the main text) to a constant value other than one. Therefore, we can say that all the relevant information can be extracted from Fig. 2 in the main text, although Fig. \ref{fig:APC} is still a good complement.

\section{F. Coexistence of discontinuous and continuous phase transitions}

\begin{figure}[H]
    \centering
    \includegraphics[width=0.35\linewidth]{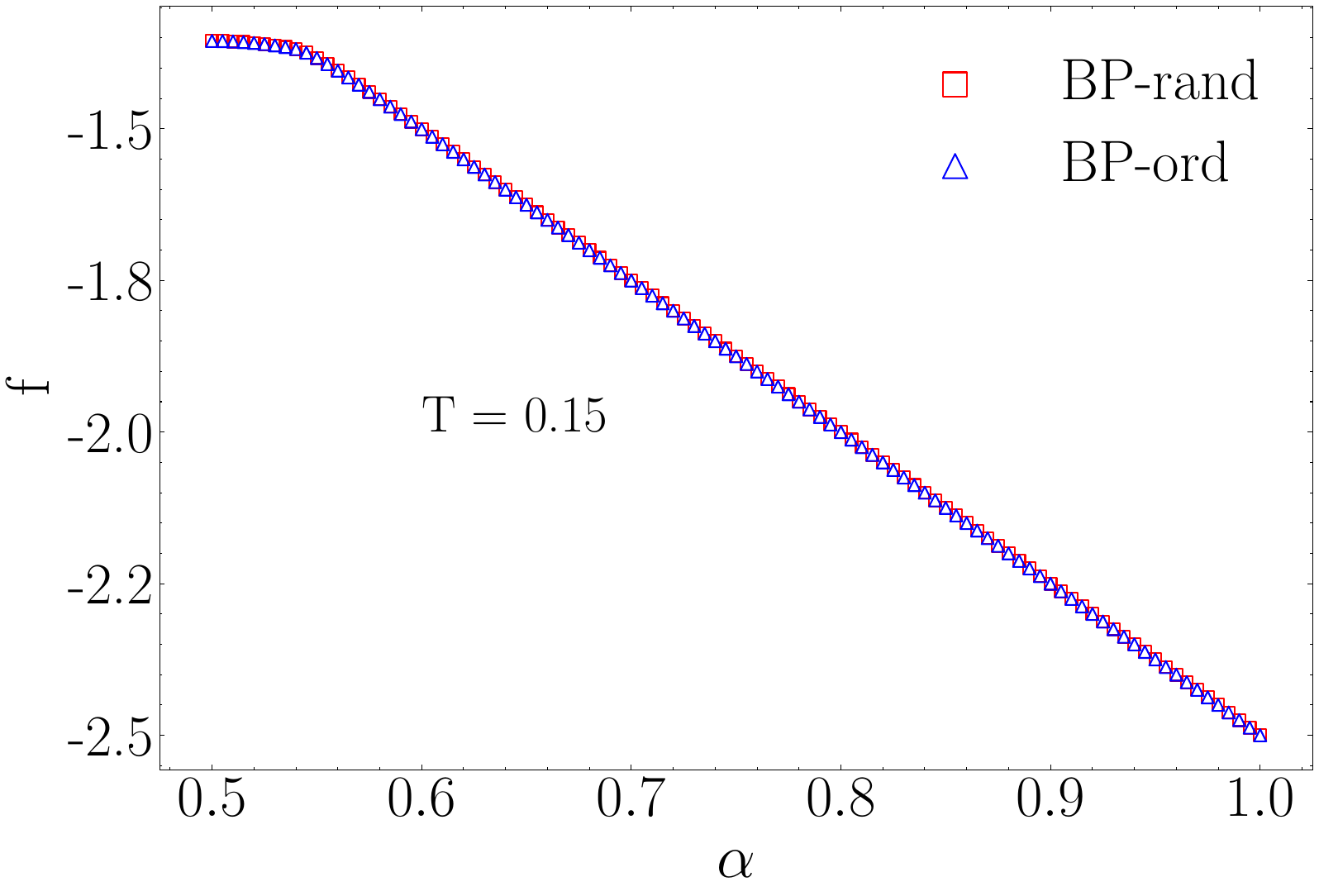}%
    \qquad \includegraphics[width=0.35\linewidth]{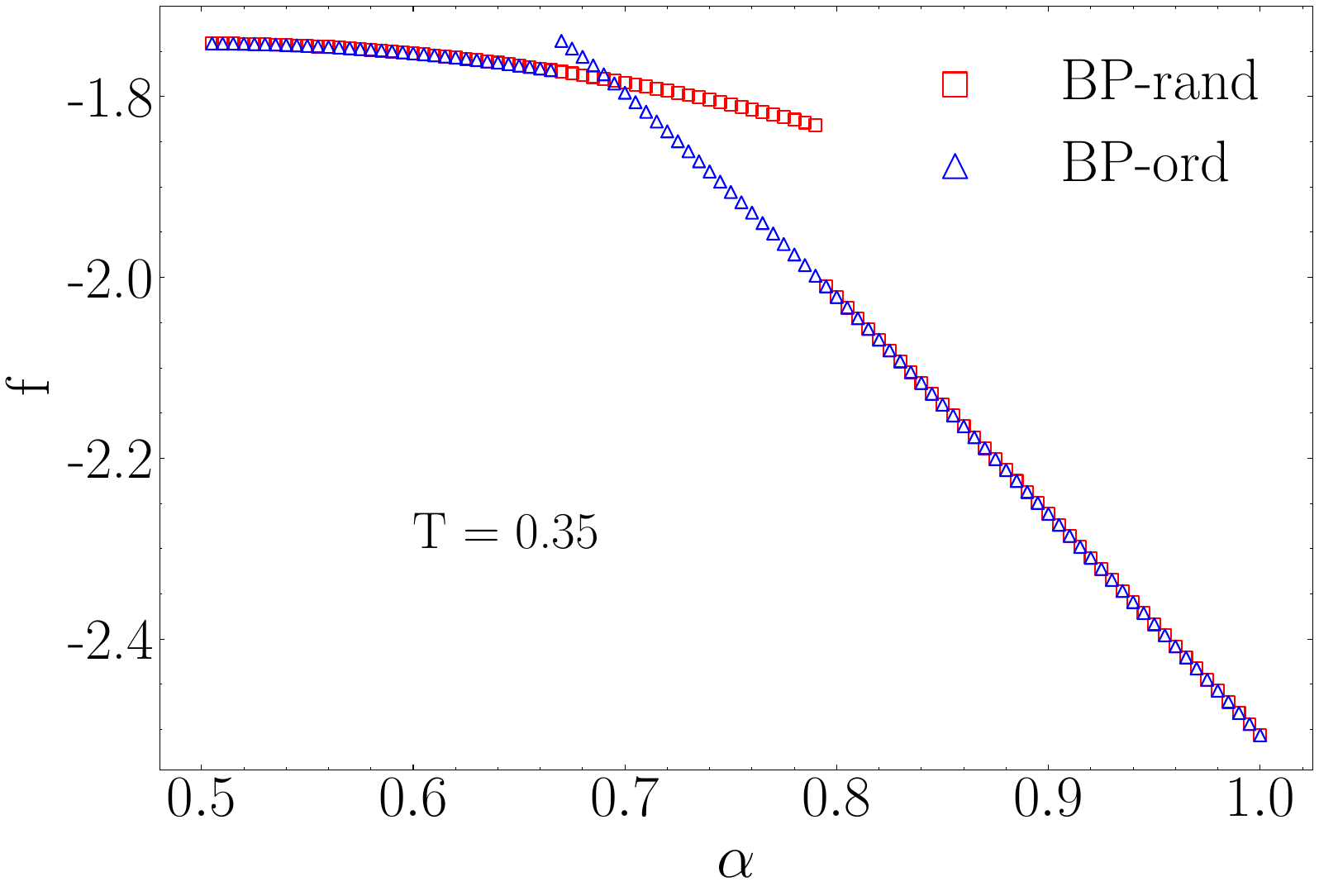}

    \includegraphics[width=0.35\linewidth]{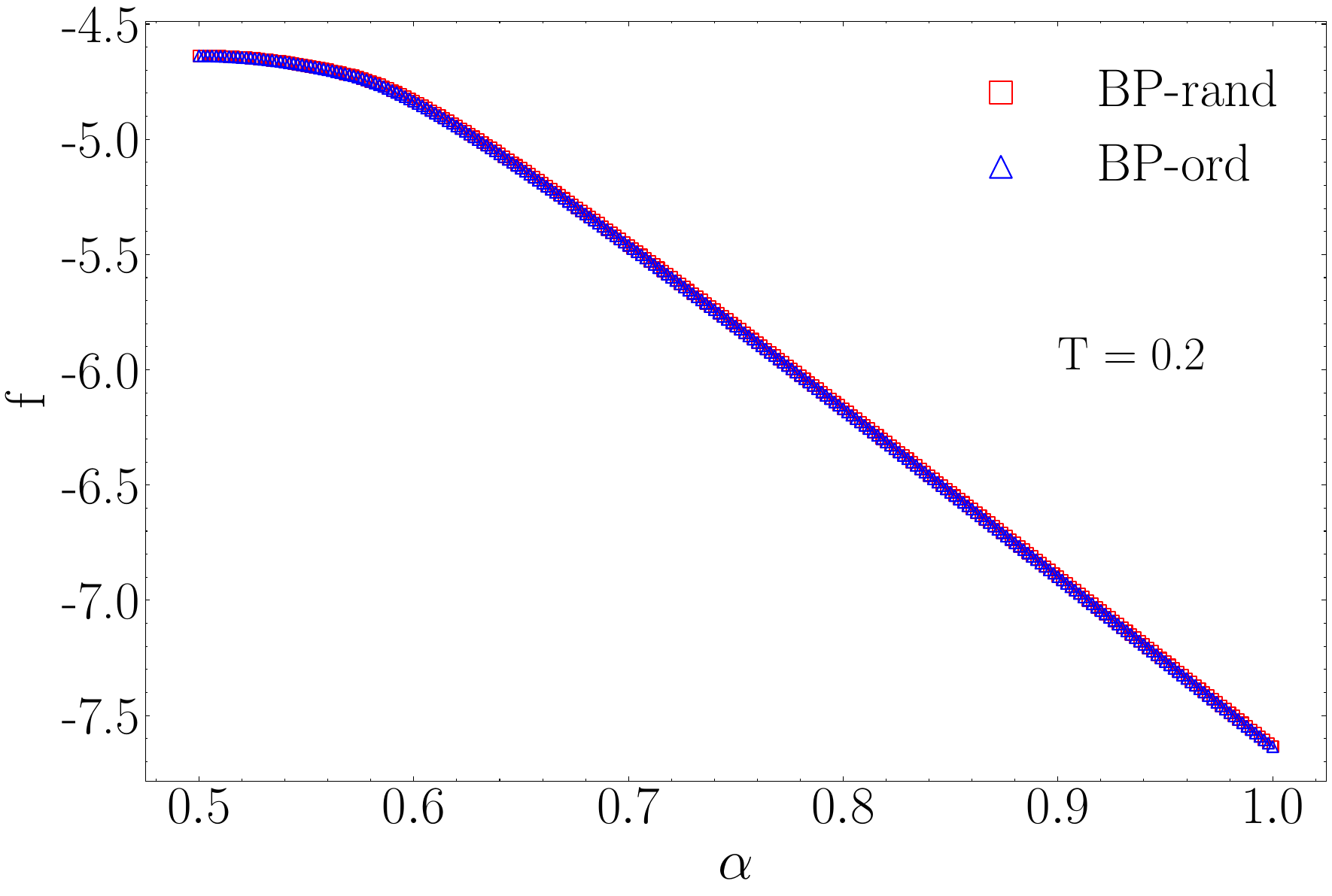}%
        \qquad \includegraphics[width=0.35\linewidth]{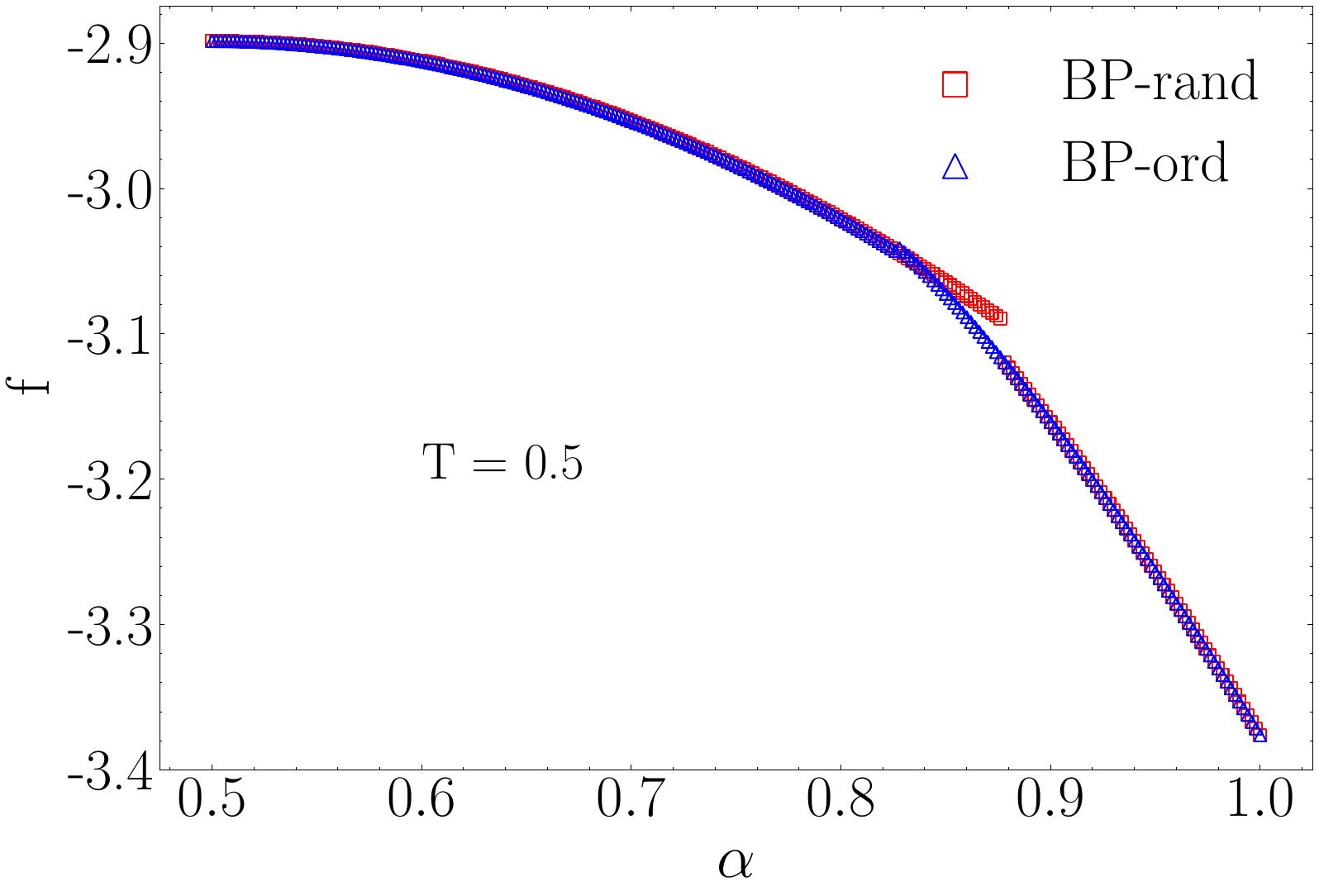}
    \caption{Free energy density computed with BP \textit{vs.} $\alpha$ for random regular (top panels) and Erdős-Rényi (bottom panels) graphs. Red squares are obtained with random initial conditions, while blue triangles are obtained with fully aligned initial conditions. The temperatures are $T=0.15$ (top-left) and $T=0.35$ (top-right) for random regular graphs. For Erdős-Rényi graphs we selected $T=0.2$ (bottom-left) and $T=0.5$ (bottom-right).}
    \label{fig:free_energy}
\end{figure}

The model defined in the main text displays several interesting features. One of them is the coexistence, for the same number of discussed issues $g$ and the same random graph of interactions, of continuous and discontinuous phase transitions. For high temperatures, the order-disorder transition (ferromagnetic state to paramagnetic state) is discontinuous. For low temperatures, instead, it is continuous. 

Fig. \ref{fig:free_energy} shows that this phenomenon is qualitatively the same in both families of random graphs: random regular and Erdős-Rényi. For low temperatures (left panels), the free energy density $f$ computed with BP is indeed a continuously differentiable function of $\alpha$. For high temperatures (right panels), we observe two distinct branches of the free energy. When two possible values coexist, the true $f$ is always the minimum. The resulting function has a discontinuity in its derivative. Thus, the transition is discontinuous.

\end{document}